\newcommand{\bb}[1]{\ensuremath{\mathbb{#1}}}
\documentclass[tightenlines,twoside,twocolumn,floatfix,nofootinbib,11pt,showpacs,pre]{revtex4-1}
\usepackage[dvips]{graphicx}
\graphicspath{{../orientationfigs/}}

\usepackage{amsmath}   % Or whichever packages you want
\usepackage{amssymb}
\usepackage{verbatim}
\usepackage[usenames,dvipsnames]{color}
\usepackage[colorlinks=true,linkcolor=blue]{hyperref}

\begin{document}

\newcommand{\TT}{\ensuremath{\mathbb{T}}}
%\DeclareGraphicsExtensions{.pdf,.png,.gif,.jpg,.jpeg,.eps}
\author{Brian Moths} 
\email{bmoths@uchicago.edu} 
\author{T. A. Witten} 
\affiliation{Department of Physics and James Franck Institute, University of Chicago, Chicago, Illinois 60637, USA} 

\title{Orientational ordering of colloidal dispersions by application of time dependent external forces}

\date{\today}

\begin{abstract}
 { We present a method of organizing incoherent motion of a colloidal suspension to produce synchronized, coherent motion.
  This method exploits general features of rotational response to time-dependent forcing, and it does not require interaction between the particles.
  We report two methods of achieving orientational alignment of an ensemble of identical colloids by means of a time-dependent, but spatially uniform forcing: a) a piecewise constant force alternating between two directions and b) a force uniformly rotating about an axis.
  The physical origin of the forcing may be e.g., sedimentation or electrophoresis. We will demonstrate that these forcing methods achieve alignment both by analyzing the equations of motion and by simulation.
  We find the conditions guaranteeing alignment, discuss the limitations of these methods, and suggest possible applications.
  %In particular, it is necessary that the colloid be asymmetric, and be at least micron size to mitigate thermal diffusion.
  %On the other hand, it is necessary that the object be small enough that its dynamics are described by low Reynolds numbers, so that an object angular velocity depends linearly on the applied force.
Examples of such forcing include electrophoresis and sedimentation. }
\end{abstract}

\maketitle

\section{Introduction}
\label{sec:Intro}

	Phase coherence plays an major role in many applications.
For example, in nuclear magnetic resonance, orientational alignment of the spins produces a large magnetic signal which is used as a characterization tool.
It is essential for lasers, where the coherence of the emitted photons imbues the light with remarkable properties.
In this paper, we will describe a method of inducing coherence in the context of colloidal sedimentation.
Specifically, we will describe a method of orientationally aligning an ensemble of colloids by uniformly applying a time-dependent force.
This alignment is desirable since once the ensemble is aligned, any response to subsequent forcing will automatically be coherent.
Thus it is possible to manipulate the orientation of objects in the ensemble in a controlled way.
	
	Our interest in this example of coherence has grown out of a body of recent work on the subject of sedimentation of colloidal objects \cite{Krapf09, Gonzalez04, Makino03,Makino05}.
These references discuss the following situation: an object with arbitrary shape is immersed in an infinite volume of fluid, which is sufficiently viscous to make the Reynolds number small.
Hydrodynamic as well as external forces (e. g., gravity) are assumed to act on the object, and the resulting positional and orientational evolution of the object is determined.
An object often exhibits a simple response to constant forcing: a direction in the object aligns with the force, and the object subsequently rotates about this axis with a constant angular velocity \cite{Gonzalez04}.
We call an object with this behavior ``axially aligning." If an entire ensemble of such objects were subject to a force, they would all become aligned with a common axis.
Still, the objects' orientations are not identical.
They differ from one another by a rotation about the force axis.
	
	The results \cite{Krapf09, Gonzalez04, Makino03,Makino05} have paved the way for further research.
For example, when the forcing is electrophoretic, \cite{Long98} explains how to generate two novel responses to the applied electric field: rotation without translation and translation perpendicular to the applied field.
Evidently, the direction of rotation depends on the handedness of the object.
This property enables one to separate enantiomers in a shear flow \cite{Doi05,Andreev10,Makino05b,Eichhorn10,Marcos09,Makino08}.
However, the problem of removing the orientational indeterminacy present after constant forcing is, to our knowledge, unaddressed.
	
	In this paper our goal is to show how to effect orientational alignment in an ensemble of colloids using time-dependent forcing.
We will study two different examples of forcing, and show, by analysis and simulation, that these lead to alignment under the suitable hypotheses.
In Section \ref{sec:formalism}, we will build the mathematical framework required to study the problem.
In Section \ref{sec:conForce} we will review the response to constant forcing.
In Section \ref{sec:compalign}, we will apply our formalism to treat two forcing programs and show that they both have the potential to align.
In Section \ref{sec:numsim}, we will present results of numerical simulations of the two forcing programs.
In Section \ref{sec:disc}, we will discuss the assumptions necessary to obtain our results, determine the minimum object size necessary to suppress thermal rotational diffusion, and briefly suggest possible applications.
	
\section{Mathematical Formalism}
\label{sec:formalism}
\subsection{Rotational Response}
\label{subsec:rotres}
We may use the established linear-response formalism to describe the orientation effects of interest \cite{Happel65}.
We are interested in the motion of an ensemble of identical objects which are dispersed in a fluid and subject to an external time-dependent force $\vec{F}_e(t)$.
For definiteness, we assume the external force is gravity; we will discuss generalizations to other types of forcing at the end of this section.
We consider the regime of ``creeping flow" \cite{Happel65}, in which inertial forces are negligible and the force transmitted to a moving object by the medium is proportional to the object's velocity.
For a \textit{rotating} rigid body, the transmitted force $\vec{F_h}$ and torque $\vec{\tau_h}$ both depend linearly on the center of mass velocity $\vec{v}$ and angular velocity $\vec{\omega}$.

How do controllable forces influence an objects orientation?
Since inertial forces are negligible, any external force must be exactly opposed by a hydrodynamic force.
For an object to generate this hydrodynamic force, it must have the appropriate velocity and angular velocity.
The orientation may be altered by prescribing the angular velocity.
Thus the external force can be used to control the orientation.
Below we look at each step in more detail.
	
	As noted above, the hydrodynamic force must satisfy $\vec{F}_h=-\vec{F}_e$.
Since \textit{rotational} inertia is also negligible, the hydrodynamic and external torques must be related similarly: $\vec{\tau}_h=-\vec{\tau}_e$.
Thus the hydrodynamic forces and torques are easily controlled.
Because these relationships are so simple, it is convenient to refer only to the hydrodynamic force and torque, and to drop their subscripts, referring to them simply as $\vec{F}$ and $\vec{\tau}$ respectively.
	
The connection between the hydrodynamic force and the object's motion is given by its linear-response properties.
To discuss these properties, it is convenient to introduce some new notation.
Let $R$ be a characteristic size of the object under consideration; for definiteness we take the stokes radius \cite{Happel65}.
Following \cite{Krapf09}, we define two six-component vectors: the generalized hydrodynamic force vector $\vec{\mathcal{F}} \equiv (\vec{F}, \vec{\tau}/R)^{T}$ specifying the force and torque, and the generalized velocity vector $\vec{\mathcal{V}} \equiv (\vec{V}, \vec{\omega}R)^{T}$ specifying the velocity and angular velocity.
We noted that it is possible to impose a desired $\mathcal{\vec{F}}$ on the system.
The object is obliged to assume the $\mathcal{\vec{V}}$ which generates this force.
In the creeping flow regime, an object moving with generalized velocity $\vec{\mathcal{V}}$ generates a generalized hydrodynamic force given by
\begin{equation}
 	\label{eq:propulsion}
 	\vec{\mathcal{F}}=6 \pi \eta R \underline{\mathbb{P}}\vec{\mathcal{V}},
\end{equation}
where $\underline{\mathbb{P}}$ is a dimensionless symmetric $6 \times 6$ matrix.
To simplify (\ref{eq:propulsion}), we choose units where $R=6 \pi \eta=1$.
	Since we seek the motional response required to generate our imposed force, we will consider the inverse of $\underline{\mathbb{P}}$, denoted $\underline{\mathbb{M}}$ \cite{Happel65}; the required $\mathcal{\vec{V}}$ is
\begin{equation}
 	\label{eq:mobility}
 	\vec{\mathcal{V}}=\underline{\mathbb{M}}\vec{\mathcal{F}}.
\end{equation}

To separate the roles of force and torque, we distinguish the four $3 \times 3$ submatrices of $\underline{\mathbb{M}}$ that give the response to $\vec{v}$ and $\vec{\omega}$ to $\vec{F}$ and $\vec{\tau}$ by writing $\underline{\mathbb{M}}$ the following way:
\begin{equation}
 	\label{eq:block}
 	\underline{\mathbb{M}}=\left( \begin{array}{cc} 	\mathbb{A} &
  \mathbb{T}^{T} \\
  \mathbb{T} &
  \mathbb{S} \end{array}	\right).
\end{equation}
In sedimentation, the external force may be regarded as acting at a single point: the center of buoyant mass.
We choose this forcing point as the origin, so that the external torque on the body is zero.
Then the rotational motion of the body is completely determined by the $3 \times 3$ ``twist matrix" $\bb{T}$:
\begin{equation}
 	\label{eq:oTF}
 	\vec{\omega}=\mathbb{T}\vec{F}.
\end{equation}
This equation captures how an applied force causes the orientation to change; it will be fundamental in what follows.
The complexity of this equation lies in the fact that the tensor $\mathbb{T}$, being a property of the object, changes when the object is rotated.
Thus, even if the force is constant, the angular velocity causes \TT{} to change, which in turn causes the angular velocity to change.
This interplay can lead to complicated dynamics as discussed in Section \ref{sec:conForce}.

%Our derivation of (\ref{eq:oTF}) had one step that may be exploited to gain control over the rotation of the object: picking the origin to be the center of buoyancy.
%If it is possible to move the center of buoyancy by changing the object's mass distribution, then the choice of origin will move to the new center of buoyancy.
%If the shape of the object does not change, then the matrix elements of $\mathbb{T}$ in the new coordinate system are related to the old matrix elements by a simple transformation law.
%Using this law, it has been shown \cite{Krapf09} that moving the origin sufficiently far causes the object to become axially aligning.
The $\mathbb{T}$ matrix of an object depends on the position of the origin, and hence on the forcing point; this point may be varied without changing the object's shape e.g., by changing the mass distribution within it. A simple transformation law determines how $\mathbb{T}$ changes under a change of forcing point \cite{Krapf09}. One location in the object called the ``center of twist" makes $\mathbb{T}$ symmetric. Moving the forcing point sufficiently far from the center of twist always causes the object to be axially aligning \cite{Krapf09}.

We may extend the results of this section to forcings of non-gravitational origin.
One example is to increase buoyancy effects via centrifugation.
The only difficulty with using a centrifugal force is its non-uniformity.
Otherwise the centrifugal force is indistinguishable from a gravitational force of the same strength, so the arguments of this section are valid without modification.
Another forcing to consider is electrophoresis.
In this case, the electric field's effect on dissolved counter-ions cause the fluid to flow.
This flow may lead to exotic motion, such as translation purely perpendicular to the applied electric field or rotation without translation \cite{Teubner82}.
Nonetheless, it has been found that even in this case, the angular velocity of an object depends linearly on the applied field.
Thus, provided that \TT{} is replaced by an effective twist matrix and $\vec{F}$ is replaced by $\vec{E}$, (\ref{eq:oTF}) continues to hold.

Having made these observations, we will no longer use any property of the object besides $\mathbb{T}$, and we use no information about the forcing except the vector $\vec{F}$ (with the understanding that it may represent an electric field or even some other vector parameter).
We will simply study the differential equation (\ref{eq:oTF}) for general $\mathbb{T}$ and $\vec{F}$.
We have reduced the problem to pure mathematics.

\subsection{Equation of Motion}
\label{subsec:eom}

In this section we will derive the equation of motion for $\mathbb{T}$ implied by (\ref{eq:oTF}).
Equation (\ref{eq:oTF}) states that when a force is applied, the object will rotate with the angular velocity determined by \TT{} and $\vec{F}$.
Between the initial time $t=0$ and a time $t$, the object will have undergone a finite rotation, given by an orthogonal transformation matrix $\mathbb{R}(t)$.
When the object does rotate, the initial twist matrix $\mathbb{T}_0$ must transform into the twist matrix $\mathbb{T}$ at time $t$ according to $\mathbb{T} = \bb{R} \bb{T}_0 \bb{R}^T$.
From this equation, it is clear that the time derivative of $\mathbb{T}$ can be expressed terms of the time derivative of $\mathbb{R}$, which we now examine.
To write this derivative, we introduce new notation: for any vector $\vec{v}$, we define the ``cross product matrix" $\vec{v}^\times$ to be the one that, acting on a vector $\vec{u}$, produces the vector $\vec{v} \times \vec{u}$. 
%\begin{equation}
% 	\label{eq:cross}
% 	\forall \vec{u} \in \mathbb{R}^3, \vec{v}^\times \vec{u} = \vec{v} \times \vec{u},
%\end{equation}
The matrix entries of $\vec{v}^\times$ are given by 
\begin{equation}
 	\label{eq:crossent}
 	\left[ \vec{v}^\times \right]_{ik} = \epsilon_{ijk} v_j.
\end{equation}
Then the time derivative of $\bb{R}$ is given by $\dot{\bb{R}}=\vec{\omega}^\times \bb{R}$.
From this expression we find \begin{equation}
 	\label{eq:diffeq}
 	\begin{aligned}
  	\dot{\bb{T}}&%
  =\frac{d}{dt} \bb{R} \bb{T}_0 \bb{R}^T\\
  &%
  =\dot{\bb{R}} \bb{T}_0 \bb{R}^T + \bb{R} \bb{T}_0 \dot{\bb{R}}^T \\
  &%
  =\vec{\omega}^\times \bb{R} \bb{T}_0 \bb{R}^T + \bb{R} \bb{T}_0 (\vec{\omega}^\times {\bb{R}})^T \\
  &%
  =\vec{\omega}^\times \bb{T} - \bb{T} \vec{\omega}^\times \\
  &%
  =[\vec{\omega}^\times, \bb{T}] \\
  &%
  =[( \bb{T} \vec{F} ) ^\times, \bb{T}], \\
  	\end{aligned}
\end{equation}
where to get the last line we have used (\ref{eq:oTF}).

\section{Constant Forcing}
\label{sec:conForce}
In this section, we review the motion resulting from a constant force.
In so doing, we will find conditions on the twist matrix which are necessary and sufficient to ensure that the object is axially aligning.
This section lays the groundwork for the following discussion of non-constant forcing.

The goal, then, is to solve the differential equation (\ref{eq:diffeq}) assuming constant $\vec{F}$.
The character of these solutions depends on the matrix $\mathbb{T}$.
There are only three classes of \TT{}, each exhibiting different behavior, of which one is axial alignment.

One class consists of symmetric \TT{} matrices.
In this case there is an analogy with classical mechanics.
If we take (\ref{eq:oTF}) and replace $\vec{F}$ and $\mathbb{T}$ by the angular momentum vector $\vec{L}$ and the inverse of the inertia tensor $\mathbb{I}^{-1}$ respectively, we are left with the equation for free rotation of a body.
Let us recall the solution to this equation.
The standard approach is to view the motion in the body frame \cite{Goldstein02}.
We can describe the matrix $\mathbb{I}$ in terms of its eigenvalues $\lambda_1$, $\lambda_2$, and $\lambda_3$, and their corresponding eigenvectors $\vec{v}_1$, $\vec{v}_2$, and $\vec{v}_3$.
Since we are in the body frame, both the eigenvalues and the eigenvectors are constant.
Let us consider the generic case where the three eigenvalues are distinct, and let us label the eigenvalues so that $\lambda_1>\lambda_2>\lambda_3$.
Now the vector $\vec{L}$ is allowed to change with time, but its length must be constant, so that $\vec{L}$ is constrained to lie on a sphere.
Since $\bb{I}$ is symmetric, the energy $E \equiv \frac{1}{2} \vec{L} \bb{I}^{-1} \vec{L}$ is also constant in time.
Thus $\vec{L} (t)$ is confined to lines of constant $E$ on the sphere.
If $\vec{L}$ is close to the $\vec{v}_1$ or $\vec{v}_3$ direction, then the energy $E$ is nearly extremal and $\vec{L}$ is confined to a small neighborhood.
Thus we see the fixed points $\pm \vec{v_1}$ and $\pm \vec{v}_3$ are (neutrally) stable.
However, if $ \vec{L}$ is initially near $\pm \vec{v}_2$, it will in general move far from this vector and so the fixed point $\pm \vec{v}_2$ is unstable.
Translating this back into the language of sedimentation, we would say that the eigenvectors $\vec{v}_1$ and $\vec{v}_3$ of $\bb{T}$ are neutrally stable fixed points for the motion of $\vec{F}$ while $\vec{v}_2$ is an unstable fixed point.

Now unlike $\mathbb{I}$, $\mathbb{T}$ need not be symmetric.
If the antisymmetric part of $\bb{T}$ is sufficiently small, then the eigenvalues remain real, and one of $\pm \vec{v}_1$ and one of $\pm \vec{v}_3$ become unstable \cite{Gonzalez04}.
For this second class of \TT{}, two objects will not, in general, align with the same axis \cite{Gonzalez04}.
Thus we call an object in this class ``non-aligning".

\begin{figure*}
 \begin{tabular}{ccc}
  \includegraphics[width=45mm]{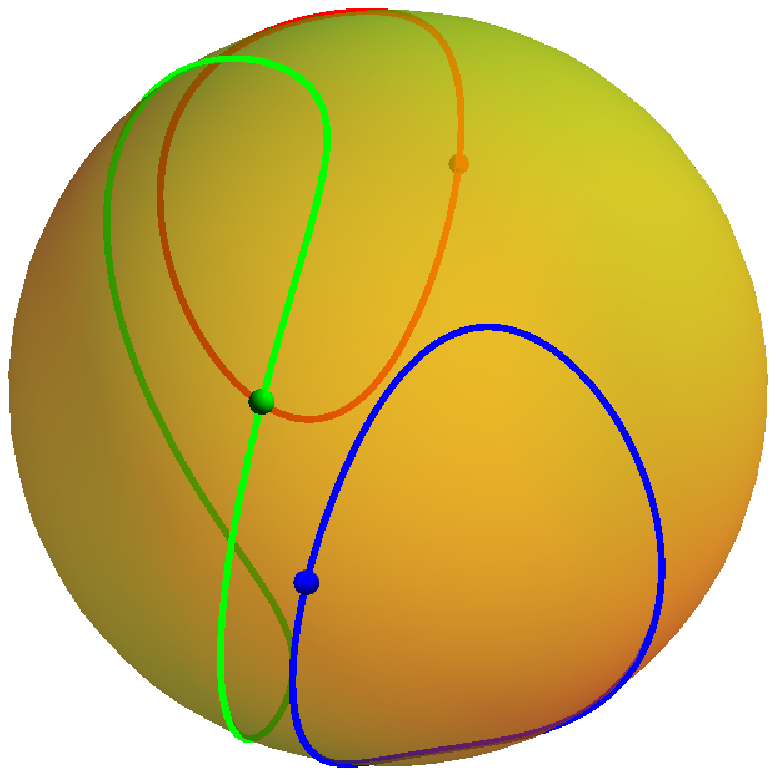} &
  \includegraphics[width=45mm]{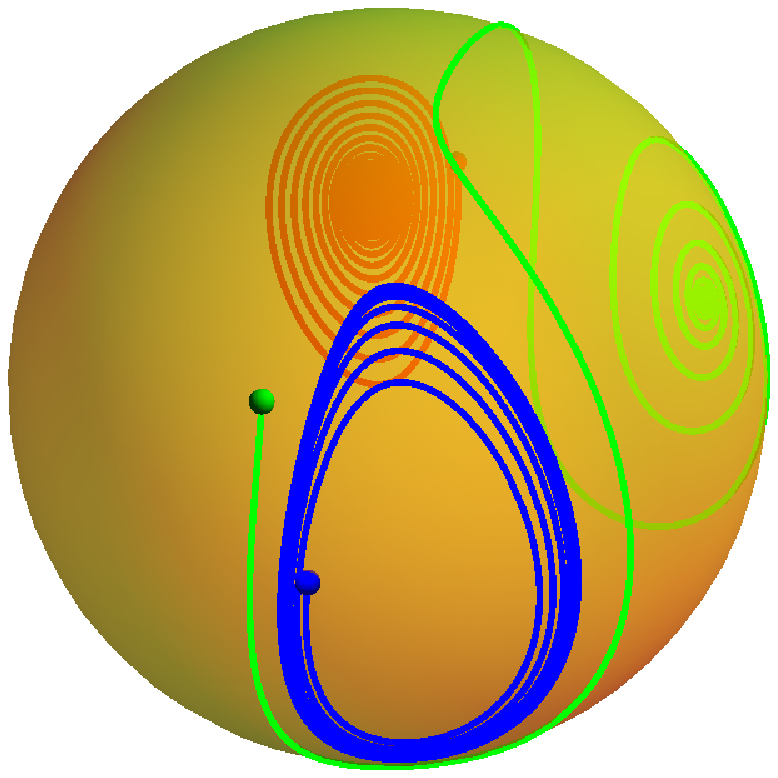} &
  \includegraphics[width=45mm]{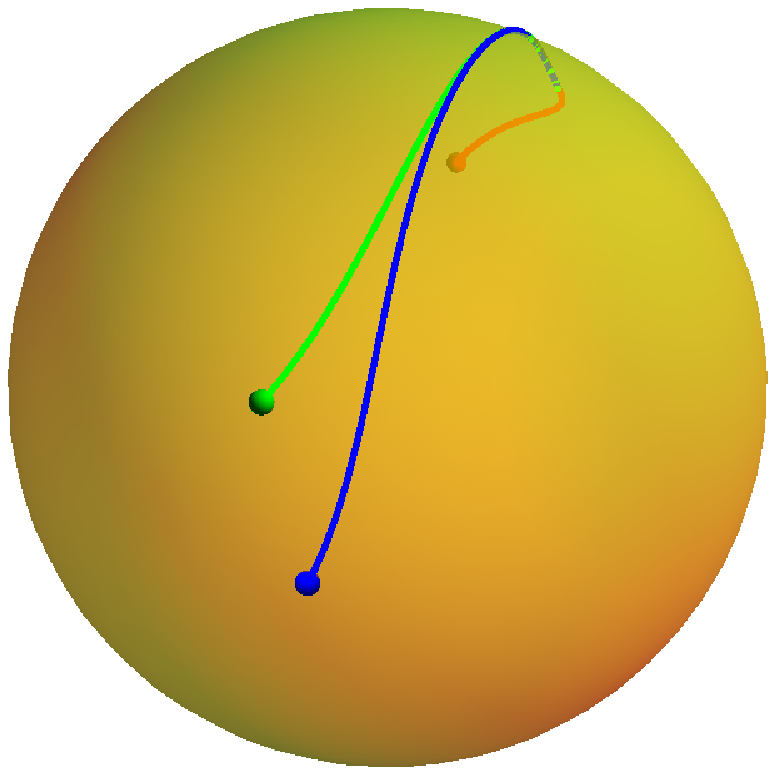}\\
  (a) &
  (b) &
  (c)
 \end{tabular}
 \caption{Illustrations of possible motions of $\vec{F}$ in the object frame for differing initial conditions and twist matrices.
  The small red, blue, and green balls show the initial value of $\vec{F}$, and the curves give the evolution of $\vec{F}$.
  The twist matrices used to generate (a), (b), (c) were symmetric, non-aligning, and axially aligning, respectively.
  In (a) the various motions of $\vec{F}$ lie on closed curves, analogously to the angular momentum in the case of a freely rotating body.
  (b) shows the effect of adding a small antisymmetric part to the \TT{} of (a).
  There are two stable fixed points, as shown by the red and green trajectories.
  The blue trajectory does not converge to a fixed point but instead converges to a limit cycle.
  (c) shows the effect of further increasing the antisymmetric part in (b) so that complex eigenvalues appear.
  Here the three point converge to a single fixed point.}
 \label{fig:flows}
\end{figure*}

Finally, a $\mathbb{T}$ may have an antisymmetric part that is too big for the second class.
This $\mathbb{T}$ belongs to the third class.
In this class, $\mathbb{T}$ has only one real eigenvalue, $\lambda_3$, the other two being complex conjugates, as is the case for a completely antisymmetric \TT{}.
For this third class of \TT{}, the motion leads to axial alignment \cite{Gonzalez04}.
In the body frame, this means that for almost any initial condition, $\vec{F}$ aligns along some direction.
Moreover, this direction is in the one dimensional $\lambda_3$ eigenspace.
This allows us to fix the sign of the eigenvector $\vec{v}_3$ so that it is the unit vector along the direction of alignment.
Then the only $\vec{F}$'s which do not align with $\vec{v}_3$ are those parallel with $-\vec{v}_3$.
Since the criterion that $\mathbb{T}$ has only one real eigenvalue is necessary and sufficient for the object be axially aligning, we will refer to these twist matrices themselves as axially aligning.
The motion of $\vec{F}$ in the body frame for the cases of symmetric, non-aligning, and axially aligning twist matrices is shown in Fig.
\ref{fig:flows}.

\begin{figure*}
 \begin{tabular}{lc}
  \raisebox{2ex}{(a)} &
  \includegraphics[width=150mm]{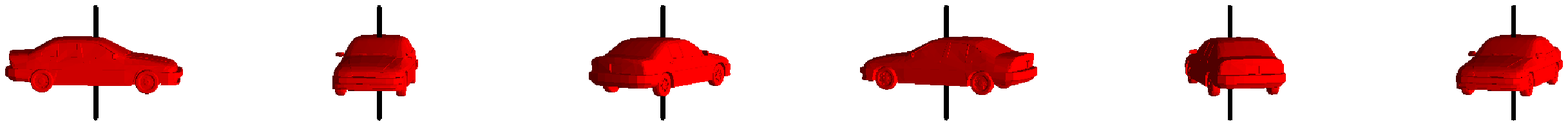} \\
  \raisebox{2ex}{(b)} &
  \includegraphics[width=150mm]{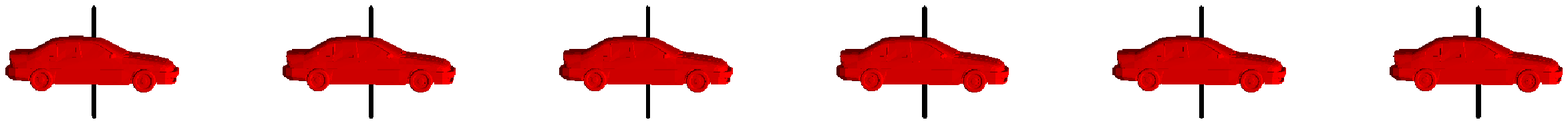}
 \end{tabular}
 \caption{Two snapshots of a set of objects with differing degrees of alignment.
  (a) Axial alignment.
  There is an axis in the object, indicated by the black line, pointing along the same direction in each instance.
  This implies that the orientation of the cars differs only by a rotation about this axis.
  (b) Each car in the ensemble is in precisely the same orientation.}
 \label{fig:axvcomp}
\end{figure*}

Let us now view the rotation of an axially aligning $\bb{T}$ in the lab frame where $\vec{F}$ is constant.
The above shows that the $\bb{T}$ matrix must rotate so that its $\vec{v}_3$ eigendirection becomes aligned with $\vec{F}$, but the orientation of the objects about $\vec{F}$ are unconstrained.
Therefore, if an entire ensemble of objects were subject to a constant force, the ensemble would become axially aligned, with each object's $\vec{v}_3$ aligned with $\vec{F}$.
We contrast this state of axial alignment with \textit{complete} alignment, where every object in the ensemble has an identical orientation.
The distinction between complete and axial alignment is illustrated in Fig.
\ref{fig:axvcomp}.
It would be convenient if complete alignment could be attained by use of a constant force.
However, since each object's $\vec{v}_3$ is aligned with $\vec{F}$, (\ref{eq:oTF}) states that each object has the same angular velocity: $\lambda_3 \vec{F}$.
Thus no change in relative orientation among objects is possible, and a time-dependent forcing program is necessary to achieve complete alignment.

\section{Complete Alignment}
\label{sec:compalign}
We are now ready to discuss the task at hand: completely aligning an ensemble of non-interacting identical objects using a time-dependent force.
That is, given a twist matrix \TT{}, we wish to find a time-dependent forcing program $\vec{F}(t)$ such that the difference between almost any two orientations goes to zero with time.
In the next paragraphs we make assumptions and general remarks, then we consider two choices of $\vec{F}(t)$.

First, the magnitude of $\vec{F}$ plays no essential role in alignment: it is evident from (\ref{eq:oTF}) that the only effect of doubling $\vec{F}$ would be to speed up the object's motion by a factor of two.
More generally, applying a force of varying magnitude is equivalent to applying a force of unit magnitude and reparameterizing time.
Since this reparameterization has no bearing on whether the final state is aligned, we may restrict our attention to forcings of unit magnitude.
Further, we non-dimensionalize the force by setting this unit equal to one.
We emphasize that the force has unit length by writing $\hat{F}$ in place of $\vec{F}$.

Next we make the assumption that \TT{} is axially aligning.
We saw in the previous section that the motion is simpler in this case, and all that needs to be done is to remove a rotational phase indeterminacy.
On the other hand, if the object were non-aligning, the response to a constant force would be more complicated: As shown in Fig.
\ref{fig:flows}, there are \textit{two} directions that may align with the force, and, moreover, alignment may not occur at all.
Clearly there is much more work to be done to achieve complete alignment in this case, and it is not as clear how to go about doing it.
Furthermore, we argued at the end of the introduction that a non-aligning object may be modified so that it becomes axially aligning.

We also make an assumption about the rotational symmetries that \TT{} possesses.
Actually, the assumption that \TT{} be axially aligning already puts restrictions on the symmetries the object can have.
Any symmetry axis of \TT{} must be an eigenvector.
Thus \TT{} may have rotational symmetries only about $\vec{v}_3$, and the most symmetry \TT{} could possibly have is circular symmetry.
However, we will assume in what follows that $\mathbb{T}$ has no rotational symmetries.
We will consider the effect of symmetry in Sec.
\ref{subsec:Assumptions}

Without loss of generality we take the real eigenvalue of \TT{}, $\lambda_3$, to be positive.
Now we are ready to discuss the two possible forcing programs.

\subsection{Step Function Forcing}
\label{subsec:cstepfn}
We begin by discussing the simplest possible time-dependent forcing program, a mere shift in the forcing direction by an angle $\theta$:

\begin{equation}
 	\label{eq:z2x}
 	\hat{F}_\theta(t) = 
 	\begin{cases}
  		\hat{x}\sin \theta + \hat{z} \cos \theta &
  \textrm{if } t< 0, \\
  \hat{z} &
  \textrm{if } t \ge 0,
  	\end{cases}
\end{equation}
where $\theta$---termed the rocking angle---is a fixed parameter which gives the angle between the initial force and the final force.
We have chosen the axes such that the force rotates onto the $z$-axis at time zero.
In this subsection, we will show that this forcing program, if repeated sufficiently with an appropriate choice of $\theta$, causes an ensemble to become completely aligned.

Let us briefly describe the argument we will present for this claim.
We start with an ensemble axially aligned with a force, and we switch the direction of the force.
After we do this, the ensemble will exhibit a transient response where it aligns with the new axis.
After this transient has passed each object simply rotates around the new force at constant angular velocity.
Since, as stated above, no alignment can occur once the steady state motion begins, we are interested only in the transient motion.
This motion causes some orientations to bunch together and others to move further apart, but we will show that on average the ensemble becomes more ordered.
Thus after many iterations complete alignment is achieved.

This subsection is organized in the following way: first, we introduce a formalism to represent the effect of the transient motion on an individual object.
Next we introduce a way of representing an axially aligned ensemble, and we determine the effect of the transient motion on the probability distribution of orientations.
Then we introduce a way to quantify the disorder in the ensemble, and we show the transient motion decreases this disorder on average, and we show that this means the ensemble must become ordered.

Let us discuss the effect of the transient.
For the sake of concreteness, we will temporarily consider the parameter $\theta$ to be fixed.
Since the transient motion may be complicated, our strategy will be to think of it as a black box: it accepts as input an object aligned with the initial force, and outputs an object aligned with the final force.
Since the dynamics are deterministic, any two objects starting with a given orientation must end up in same orientation.
Thus the black box defines a function from the space of initial orientations to the space of final orientations.
Since the objects must initially be aligned with $\hat{F}$, they have only one orientational degree of freedom, and the space of initial orientations is the unit circle, denoted $S^1$.
Similarly the space of final orientations is also $S^1$.
To represent the input orientation, we first arbitrarily pick a member of the ensemble as a reference object.
Then any object in the ensemble can be realized by rotating the reference object by an angle $\phi$ about $\hat{F}$.
We use this angle $\phi$ to refer to the initial orientation.
We can analogously represent the output orientation, which we will call $\psi$.
For consistency, we pick the reference object for $\psi$ to be the final orientation of the initial reference object.
With this notation in place, we can now represent the effect of the transient by the function $\psi(\phi)$.
We note that our choice of initial and final reference objects is summarized by the relation $\psi(0)=0$.
If we allow the angle $\theta$ to be chosen freely again, then the function $\psi(\phi)$ will depend on the parameter $\theta$.
We indicate this dependence by writing the function as $\psi_\theta$.
Also, it will be convenient to require that the choice of initial reference orientation be smooth in $\theta$.

For our purposes, we could forget everything about the object, including $\mathbb{T}$, except for $\psi_\theta$.
A few general properties of $\psi_\theta$ will prove useful.
First, $\psi_0$ is the identity function, since $\theta=0$ corresponds to a constant force, so each orientation advances by the same amount. Because the choice $\psi_\theta (0)=0$, this amount must be a multiple of $2\pi$.
This trivial behavior occurring when $\theta=0$ cannot be used for alignment.

The simple behavior of $\psi_0(\phi)$ leads to simple behavior for small, non-zero $\theta$.
First, $\psi_\theta$ is monotonic for sufficiently small $\theta$.
To prove this statement, we first give the reason why, in (\ref{eq:z2x}), we chose the initial force to depend on $\theta$ instead of the final force.
We made this choice because we wanted the combination $(\theta,\phi)$ to represent an initial condition for (\ref{eq:diffeq}): the object with orientation $\phi$ when the rocking angle is $\theta$.
This initial condition is smooth in $(\theta,\phi)$, because we chose the reference orientation to depend smoothly on $\theta$.
Then since the solution of a differential equation depends smoothly on initial condition \cite{Arnold92}, $\psi_\theta(\phi)$ depends smoothly on $(\theta,\phi)$.
Then so does the $\phi$-derivative.
Since this derivative is identically $1$ for $\theta=0$, it must be positive everywhere for sufficiently small $\theta$.
However, for certain objects and large enough $\theta$ it may be that $\psi_\theta$ is non-monotonic.

\begin{figure*}
 \includegraphics[width=150mm]{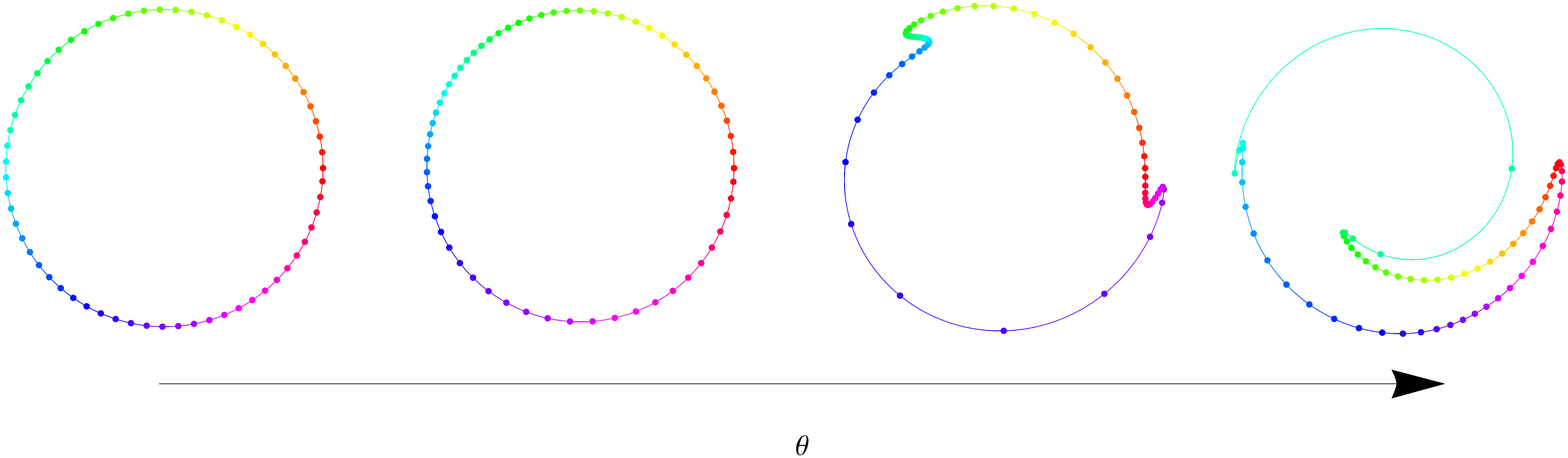}
 \caption{Illustration of the proof that $\psi_\theta$ wraps once around $S^1$.
  Four graphical representations of $\psi_\theta$ are shown for four different values of the rocking angle $\theta$.
  The color of a point on the curve indicates the initial angle $\phi$, and the azimuthal coordinate of a point on the curve is equal to $\psi_\theta(\phi)$.
  The radial coordinate has no physical significance; its purpose is to avoid self-intersections of the curve.
  For the far left figure, $\theta=0$, so that $\psi_\theta$ is the identity function.
  As $\theta$ is increased from 0, the azimuthal coordinate of each point changes, deforming the curve.
  However, since this deformation is smooth, it is impossible to change how many times $\psi_\theta$ winds around $S^1$ just by changing $\theta$.}
 \label{fig:tauwrap}
\end{figure*}

Our second result is that as the input orientation $\phi$ is increased from $0$ to $2 \pi$, the output orientation $\psi_\theta(\phi)$ undergoes a net increase of $2 \pi$ as well.
This is the proper extension of what we found for $\theta=0$; it holds for \textit{any} $\theta$.
Mathematically, this assertion can be written \begin{equation}
 	\label{eq:wrap}
 	\oint_{S^1} {{d \psi_\theta} \over {d \phi}} \, d \phi = 2 \pi, \end{equation} Since $\psi_0$ is the identity function, (\ref{eq:wrap}) is clear in the case $\theta=0$.
Our strategy to prove (\ref{eq:wrap}) in the general case $\theta \neq 0$ is to continuously connect this situation with the constant force situation.
Since the solution of a differential equation depends continuously on initial condition, the function $\psi_\theta$ can be deformed into $\psi_0$ simply by decreasing $\theta$.
Therefore these two functions must wind the same number of times around $S^1$.
From the $\theta=0$ case, we see that it must wind exactly once.
This proof of (\ref{eq:wrap}) is illustrated in Fig \ref{fig:tauwrap}.
This concludes our discussion of how the transient acts on individual objects.

Now we describe how the transient acts on the ensemble as a whole.
We will characterize an axially aligned ensemble by a probability density function (pdf) $p(\phi)$ which gives the probability of a randomly selected object having the orientation $\phi$.
Complete alignment means that this pdf is a delta function.

We start by considering an operation more simple than $\psi_\theta$.
We will study a uniform shift in orientation (henceforth, a ``shift"), which can be effected by allowing an axially aligned ensemble to rotate for some amount of time.
Although, as we have already stated, a shift does not increase alignment, it is be an essential ingredient in the forcing program we propose.
The orientation $\tilde{\phi}$ after the shift depends on the orientation $\phi$ before the shift as follows: $\tilde{\phi} = \phi + \alpha$.
The new pdf is then given by $\tilde{p}(\tilde{\phi})=p(\phi)$; since the new pdf is essentially identical to the old pdf, we see that no ordering has been achieved.

\begin{figure*}
 \begin{tabular}{cc}
  \includegraphics[width=70mm]{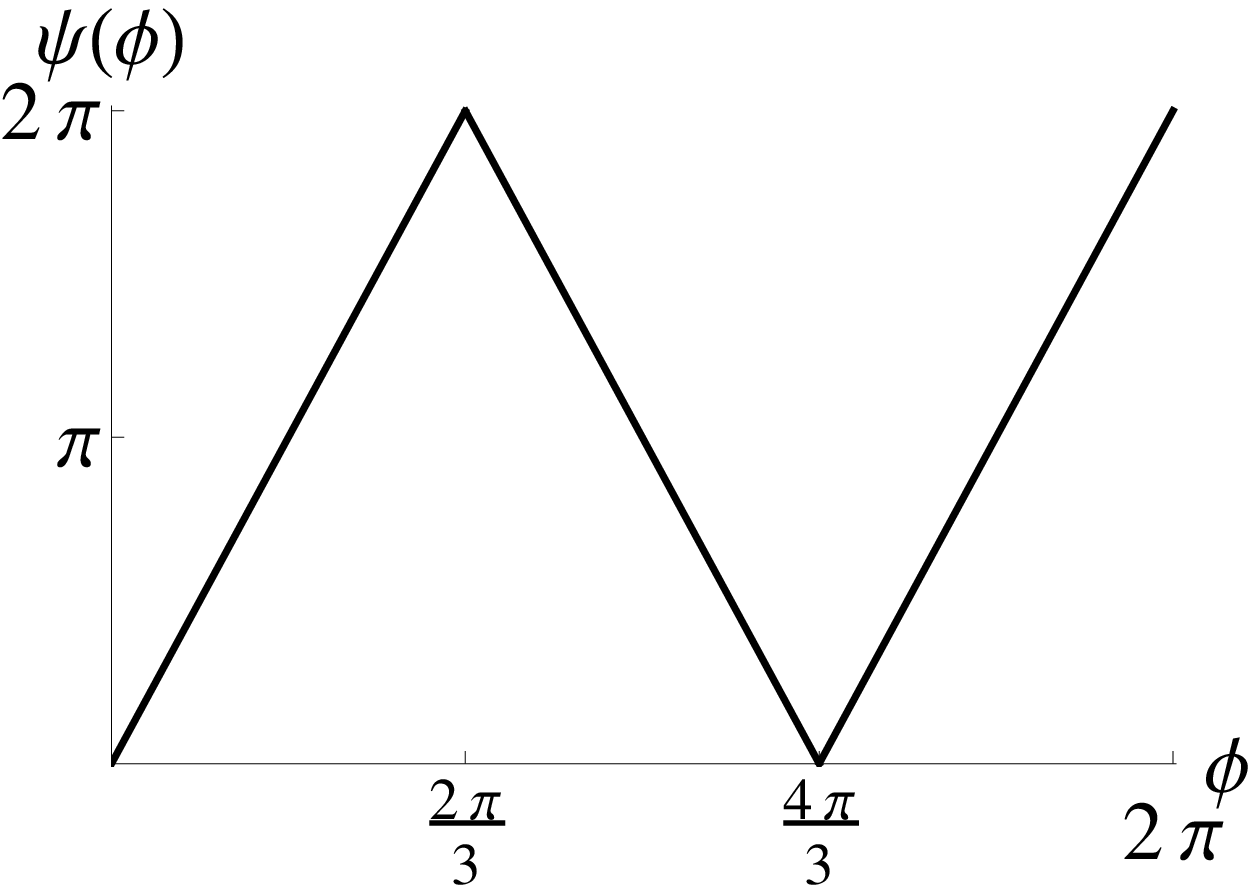} &
  \includegraphics[width=70mm]{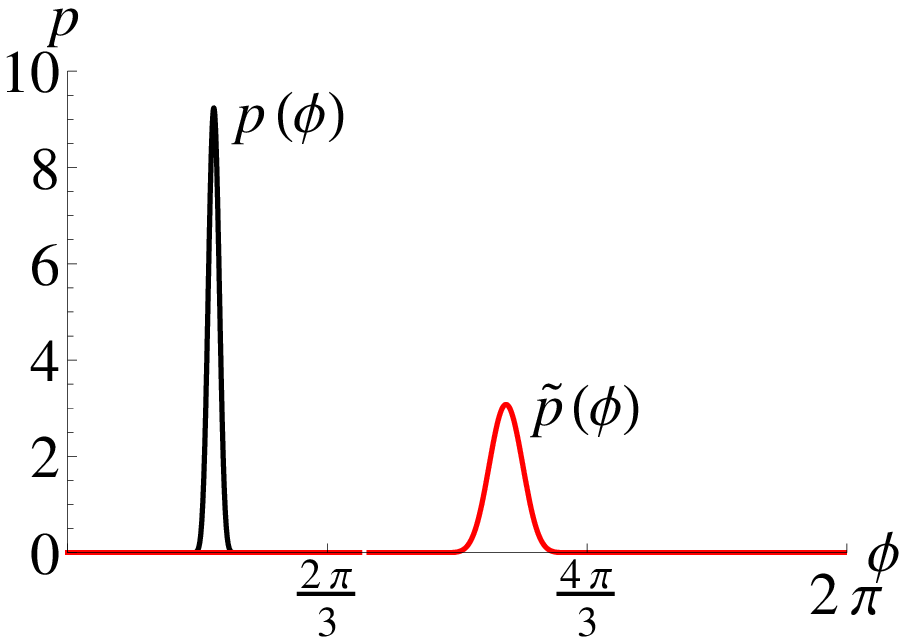}\\
  (a) &
  (b)
 \end{tabular}
 \caption{(a) Plot of $\psi(\phi)$ vs $\phi$ for a non-monotonic $\psi$ that cannot possibly lead to ordering.
  (b) A plot of a narrowly peaked distribution $p(\phi)$, shown in black, together with $\tilde{p}(\phi)$, the result of acting by $\psi$ on $p(\phi)$, shown in red.
  $\tilde{p}$ is broader and therefore represents a less ordered distribution.}
 \label{fig:badtau}
\end{figure*}

Now let's consider $\psi_{\theta}$, the effect of a single shift in forcing.
It is convenient to assume that the derivative of $\psi_\theta$ is everywhere positive, which can be assured, as explained above, by making $\theta$ sufficiently small.
We make this assumption for two reasons: it excludes certain pathological $\psi_\theta$'s, and it simplifies the transformation law of the pdf.
To help motivate this assumption, we consider a pathological $\psi_\theta$;it is shown in Fig.
\ref{fig:badtau} (a).
Note that while this $\psi_\theta$ is a valid example insofar as it satisfies the winding number constraint, it is nevertheless clear that no ordering can be achieved, because any localized distribution will be broadened by the action of $\psi_\theta$, as shown in Fig.
\ref{fig:badtau} (b).
The assumption that $\psi_\theta$ is monotonic, together with the condition that $\psi_\theta$ winds only once around $S^1$, guarantees that there is a limit to how much stretching $\psi_\theta$ can do.

Now we are ready to examine how the pdf transforms under the action of $\psi$ (we no longer concern ourselves with the value of the parameter $\theta$ and we write simply $\psi$ in place of $\psi_\theta$).
In fact, it is no more complicated to find how the pdf transforms under the composition of $\psi$ with a shift.
The orientation after this operation is \begin{equation}
 	\label{eq:tildephi}
 	\tilde{\phi} \equiv \psi(\phi + \alpha), \end{equation} where $\alpha$ is the the size of the shift; it gives additional freedom which may be used to help alignment.
From the rule $\tilde{p}(\tilde{\phi})d\tilde{\phi}=p(\phi)d\phi$, we find that the pdf transforms according to the following formula: \begin{equation}
 	\label{eq:pdf}
 	\tilde{p}(\tilde{\phi})=\frac{p(\phi)} {\psi ' (\phi+\alpha)}, \end{equation} where $\phi = \psi^{-1}(\tilde{\phi}) - \alpha$.

Now that we know how the ensemble can be manipulated, we must ask if it is becoming more or less ordered.
We will use the information-theoretic entropy, $H$ \cite{Shannon_1948}, to quantify the disorder.
Given a pdf, $p(\phi)$, the functional $H[p]$ is defined as \begin{equation}
 	\label{eq:entropy}
 	H[p]=-\oint_{S^1} p(\phi) \ln{p(\phi)} \, d \phi.
\end{equation}
To justify our use of the entropy, let us recall a few of its properties.
First, a change in scale in $\phi$, say from radians to degrees, causes a change in $H$ by an additive constant.
In this sense, the zero of entropy is arbitrary, similar to the case of energy.
Another important property is that it is maximal for a uniformly distributed variable.
This means there is an upper bound on the entropy.
However there is no lower bound.
In fact a pdf limited to some region of size $\Delta \phi$ can have an entropy of at most $\log (\Delta \phi)$.
This upper limit makes it clear that the entropy goes to $- \infty$ as the distribution becomes more and more narrow.
These observations lends credence to the interpretation of $H$ as a measure of disorder.

We now calculate the change in entropy resulting from the transient.
The entropy $\tilde{H}_\alpha$ after the transient with shift $\alpha$ is given by $\tilde{H}_\alpha=-\oint_{S^1} \tilde{p}(\tilde{\phi}) \ln (\tilde{p}(\tilde{\phi})) \, d\tilde{\phi}$.
Upon substituting (\ref{eq:pdf}) and changing the integration variable from $\tilde{\phi}$ to $\phi$, we find $\tilde{H}=H + \Delta H_\alpha$ where 
\begin{equation}
 	\label{eq:dhpsi}
 	\Delta H_\alpha = \oint_{S^1} p(\phi) \ln \left( \psi ' (\phi+\alpha)\right) \, d \phi.
\end{equation}

If the new distribution is to be more ordered than the old distribution, $\Delta H_\alpha$ must be less than zero.
%
%However, $\Delta H_\alpha$ may be positive, for consider a pdf which is highly peaked about some orientation $\phi_0$ (so that it is nearly a delta function).
%Then, from (\ref{eq:dhpsi}), the change in entropy is nearly $\ln(\psi'(\phi_0+\alpha))$ which can be made positive or negative since $\psi'$ has to take on values both greater and less than unity (unless $\psi$ is the identity function).
%
However, $\Delta H_\alpha$ is certainly greater than zero for some choices of $p(\phi)$ and $\alpha$. For example, if $p(\phi)$ is strongly concentrated in a region where $\psi'>1$, then $\Delta H_0$ is positive.

Even though the change in entropy is not negative for all values of $\alpha$, we now show that if $\alpha$ is chosen randomly, the expected entropy change will be negative.
The $\alpha$-averaged change in entropy $\langle \Delta H_\alpha \rangle$ is given by ${1 \over 2 \pi} \oint_{S^1} \Delta H_\alpha \, d \alpha$.
Using (\ref{eq:dhpsi}) for $\Delta H_\alpha$, we find $\langle \Delta H_\alpha \rangle = {1 \over 2 \pi} \oint p(\phi) \, d \phi \oint \ln (\psi'(u)) \, du$.
The integral over $\phi$ is unity, so any dependence of the average entropy change on the initial pdf disappears.
Also, because the log is a concave function, $\frac{1}{2 \pi} \oint \ln (\psi'(u)) \, du \le \ln (\oint \frac{1}{2 \pi} \psi'(u) \, du)=0$, with equality only in the case that $\psi$ is the identity.
Thus we conclude $\langle \Delta H_\alpha \rangle$ is negative and independent of the initial probability distribution $p$.

%Now there is one more subtlety to be discussed.
%What we have shown is that given some initial $p$, the decrease in entropy in a single iteration, averaged over the random phase shift $\alpha$, is zero.
%What one would actually do in experiment is apply many iterations and average the entropy decrease over iteration number.
%\textit{This} is what we want to show is negative, so that the entropy decreases indefinitely as the forcing program is iterated.
%Notice two things: (1) the average over iteration number is still a random variable because the shifts in each iteration have \textit{not} been averaged over, and (2) the entropy decreases for the different iterations are not identically distributed, since each iteration begins with a different pdf.
%Nonetheless, since each of these distributions has the same mean, the central limit theorem \cite{Hall80} guarantees that for a large number of iterations, the distribution of the iteration-averaged entropy decreases will be sharply peaked around the shift-averaged entropy decrease for a single iteration, and so the entropy will almost certainly decrease indefinitely.
%%(may be ergodic theorem for markov chains)

With this information about $\langle \Delta H_\alpha \rangle$ we can infer the change in $H$ after many repetitions of the rocking procedure with randomly chosen shifts. We denote the average change in $H$ after many such iterations as $\overline{\Delta H}$. The average of $\Delta H$ after $n$ iterations with a particular randomly chosen sequence of shift angles $\{\alpha_1, \alpha_2, ... , \alpha_n\}$ is given by
\begin{equation}
\label{eq:overlineDeltaH}
\overline{\Delta H} = \frac{1}{n} \sum_{i=1}^n \Delta H_{\alpha_i}[p_i]
\end{equation}
where $p_i(\phi)$ is the probability distribution before the $i$th iteration. The ensemble average of $\overline{\Delta H}$ over the (randomly chosen) $\alpha_i$ is then given by
\begin{equation}
\label{eq:expectationOverlineDeltaH}
\langle \overline{\Delta H_\alpha} \rangle =  \frac{1}{n} \sum_{i=1}^n \langle \Delta H_{\alpha_i}[p_i] \rangle
\end{equation}
Though $p_i$ in each term of this sum depends on previous $\alpha_j$'s, it is independent of $\alpha_i$. Thus each term average is given by the $\langle \Delta H_\alpha \rangle$ calculated in (\ref{eq:dhpsi}). This average is independent of the $p_i$. Thus
\begin{equation}
\label{eq:expectationOverlineDeltaHSimplified}
\langle \overline{\Delta H_\alpha} \rangle = \langle \Delta H_\alpha \rangle = \frac{1}{2 \pi} \oint \ln \psi'(u) \, du
\end{equation}
Since the expectation of the iteration average is the same as the expectation of a single iteration, $H$ is expected to decrease indefinitely after many iterations. 
Moreover, since $\psi '$ has an upper bound and is bounded below by $0$, the range of values $\Delta H_\alpha$ can take at each iteration is also bounded.
Then the central limit theorem \cite{Hall80} guarantees that $\overline{\Delta H}$ approaches $ \langle \Delta H_\alpha \rangle $ as the number of iterations $n$ goes to infinity.

Now we see why it is important to compose $\psi$ with the shift.
With no phase shift, it is difficult to rule out the possibility that upon becoming sufficiently ordered, the system would consistently lose its order at the next iteration.
However, with randomness it is impossible for the system to consistently pick shift angles that cause disorder.

The indefinitely small entropy found above does not guarantee alignment to a single orientation: there may be \textit{two} (or more) final orientations.
However, a recent argument \cite{Kaijser1993} suggests that the final state has only one.
This concludes our proof that repeatedly applying (\ref{eq:z2x}) with sufficiently small $\theta$ will cause the ensemble to become completely aligned.

\subsection{Rotating Force}
\label{subsec:cRotF}
One shortcoming of the approach described in the previous section is that the final orientation is arbitrary.
In this section we consider an alternative forcing program, in which the final orientation is controlled. In this program, the force rotates about the $z$-axis with constant angular velocity: 
\begin{equation}
 	\label{eq:fxy}
 	\hat{F}(t)=\hat{z} \cos \theta_0 + (\hat{x} \cos \Omega t + \hat{y} \sin \Omega t ) \sin \theta_0 .
\end{equation}
Here, $\theta$ and $\Omega$ are adjustable forcing parameters.
Without loss of generality, we take $\Omega>0$ and $0 \le \theta < \pi$.
Since we have not specified the direction of the z-axis, it is no more general to allow the force to rotate about an arbitrary axis.
This axis will be indicated by the vector $\vec{\Omega}$, which points in the direction of rotation of the force and has length $\Omega$.

If the ensemble becomes completely aligned, all objects will undergo an identical motion.
Since the force is not constant, this motion could be very complicated.
However, we will show in this section that for a certain choice of parameters, the ensemble does become aligned, and the subsequent motion is simple.
This subsection is organized as follows.
First, we will describe the simple motion occurring after alignment.
Next, we will examine what choice of forcing parameters gives rise to this simple motion.
Finally we will show that with this choice of forcing parameters the motion is locally stable.
Numerical studies reported in Section \ref{subsec:nstepfn} indicate that this motion is globally stable.

We first describe the simple motion which will occur after alignment.
This motion is rotation with the same constant angular velocity as $\hat{F}$: $\vec{\Omega}$.
The direction of the force relative to the object does not change, so the state of the system at a later time is just a rotated version of the initial state, and the motion persists.
Since the object and the force are rotating together, we call this motion ``co-rotation".
It is characterized by the condition $\vec{\Omega}=\mathbb{T} \hat{F}$, called the co-rotation condition.
Notice, though, that this motion may not be possible for all choices of parameters.
For example, $\Omega$ cannot be larger than the maximum of $\Vert \mathbb{T} \hat{F} \Vert$.

%Given an object with twist matrix $\mathbb{T}_0$, we seek some some choice of parameters such that regardless of initial orientation, the twist matrix becomes and remains ``aligned" with the force. Here, by ``aligned", we mean that each object maintains the same fixed orientation with respect to both the force and the axis of rotation of the force. For there to be alignment, we must have in particular that the angular velocity of each object must be the same as the angular velocity of $\hat{F}$. We will call the requirement that the object rotate with the same angular velocity as the force the ``co-rotation condition". 

We now determine explicitly what conditions on the forcing parameters are necessary for there to be a co-rotating state.
Here we shall express $F$, $\mathbb{T}$, etc. in the body frame, indicated by a subscript ``b".
We first observe that for any initial $\hat{F}_b$, there is a unique $\vec{\Omega}_b$ that gives co-rotation, namely $\vec{\Omega}_b = \vec{\omega}_b = \mathbb{T}_b \hat{F}_b$.
Next, the length of this $\vec{\Omega}_b$ and the angle it makes with the initial $\hat{F}_b$ determines forcing parameters $\Omega$ and $\theta$ compatible with this $\hat{F}_b$.
Thus each $\hat{F}_b$ generates a particular choice of the parameters $\Omega$ and $\theta$.
However, in practice, it is not the direction of the force, but the parameters $\Omega$ and $\theta$ which are directly controlled.
For a given $\Omega$ and $\theta$, there may several $\hat{F}_b$'s which generate these parameters, i.e., several co-rotating orientations of the object in the lab frame.
On the other hand, it may be the case that there are no such $\hat{F}_b$'s.
We will find a particular choice of $\Omega$ and $\theta$ such that there is certainly at least one $\hat{F}_b$ which generates the chosen $\Omega$ and $\theta$.

We begin with an object with twist matrix $\mathbb{T}_b$, and apply a force $\hat{F}_b$.
In this frame, the co-rotation condition is $\mathbb{T}_b \hat{F}_b = \vec{\Omega}_b$.
So $\vec{\Omega}_b$ is fixed by our choice of $\hat{F}_b$.
The parameter $\Omega$ is fixed by the ``$\Omega$ constraint"
\begin{equation}
 	\label{eq:normeq}
 	\Omega^2 = \vec{\Omega}_b \cdot \vec{\Omega}_b = \hat{F}_b^T \mathbb{T}_b^T \mathbb{T}_b \hat{F}_b,
\end{equation}
and $\theta$ is fixed by the ``$\theta$ constraint" 
\begin{equation}
 	\label{eq:doteq}
 	\cos \theta = \frac{\hat{F}_b \cdot \vec{\Omega}_b}{\Omega_b} = \frac{\hat{F}_b \cdot \mathbb{T} \hat{F}_b}{\Vert \mathbb{T}_b \hat{F}_b \Vert}.
\end{equation}

Now that we know what forcing parameters are generated by a particular choice of $\hat{F}_b$, we will ask what range of forcing parameters could be generated by \textit{some} $\hat{F}_b$.
First, let's consider the $\Omega$ constraint, (\ref{eq:normeq}).
Since $\mathbb{T}_b^T \mathbb{T}_b$ is a symmetric, positive definite matrix, its eigenvalues are real and positive.
Therefore, the range values that $\hat{F}_b^T \bb{T}^T_b \bb{T}_b \hat{F}_b$ takes is the interval $[\Sigma_3, \Sigma_1]$, where $\Sigma_3$ ($\Sigma_1$) is the smallest (largest) eigenvalue of $ \bb{T}^T_b \bb{T}_b$.
Evidently $\Omega^2$ must lie in the range $\Sigma_3 \le \Omega^2 \le \Sigma_1$ for co-rotation to be possible.
If some value of $\Omega$ is chosen, then $\hat{F}_b$ is constrained to lie on the corresponding level set of $\hat{F}_b^T \bb{T}^T_b \bb{T}_b \hat{F}_b$.
A level set is the intersection of the ellipsoid defined by (\ref{eq:normeq}), and the unit sphere since we took all forces to have unit length.
We will call each one of these level sets ``$\Omega$-curves".
Notice that the $\Omega$-curves are symmetric under reflection through the origin, and they are generically a pair of closed loops (hence the plural).

Let's examine how the $\Omega$-curves change as $\Omega^2$ changes.
If $\Omega^2=\Sigma_3$, then the only $\hat{F}_b$'s capable of producing such a small $\Omega$ point along the eigendirection associated with $\Sigma_3$, so that the $\Omega$-curves are simply a pair of antipodal points.
As $\Omega^2$ increases from $\Sigma_3$, the $\Omega$-curves form increasingly larger loops around ($\pm$) the eigenvector corresponding to $\Sigma_3$.
As $\Omega^2$ reaches the middle eigenvalue of $\bb{T}^T_b \bb{T}_b$, the two loops meet at ($\pm$) the corresponding eigenvector, and as $\Omega^2$ increases more, the loops shrink to ($\pm$) the eigenvector corresponding to $\Sigma_1$.
The $\Omega$-curves look similar to the orbits of the angular momentum in the body frame of a freely rotating object as discussed previously.

Let's proceed analogously with the $\theta$ constraint (\ref{eq:doteq}).
For some choice of $\theta$, we call the set of $\hat{F}_b$'s that solve this equation $\theta$-curves.
First we consider the possible range of $\theta$.
We know we can achieve $\theta=0$ by choosing $\hat{F}_b=\vec{v}_3$.
However, the other extreme, $\theta=\pi$ is unattainable since $\mathbb{T}$ has no negative eigenvalue.
So the range of values $\theta$ can take goes from zero up to some maximum, which is less than $\pi$.
Notice that, like the $\Omega$-curves, the $\theta$-curves are symmetric with respect to reflections through the origin.
Let's see how the $\theta$-curves change as $\theta$ is changed.
When $\theta=0$, the $\theta$-curves are just the pair of points $\pm \vec{v}_3$.
As $\theta$ is increased, the $\theta$-curves expand from these two points.
Below we shall consider only the regime of small $\theta$.

\begin{figure}
 \includegraphics[width=60mm]{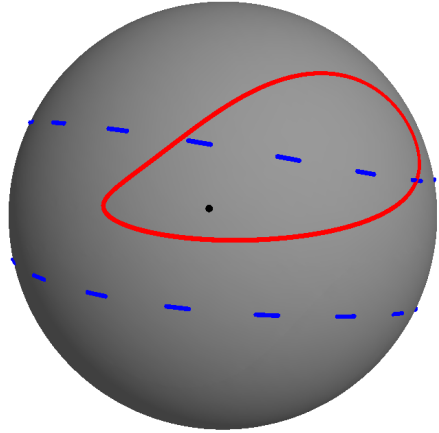}
 \caption{The $\Omega$- and $\theta$-curves.
  We have chosen an arbitrary twist matrix, and chosen $\Omega$ and $\theta$ to be compatible with co-rotation.
  This figure shows a part of each $\Omega$-curve (dashed) and one of the $\theta$-curves (solid).
  For reference, $\vec{v}_3$ is shown by a black dot.
  Sections of the two $\Omega$ curves are visible above and below the dot.
  Also, there is another $\theta$-curve antipodal to the one visible in the figure.
  We see that the there are two visible intersection points of these sets of curves, and by the inversion symmetry, the antipodes of these two intersection points are also intersection points.
  Thus there are four co-rotating states.}
 \label{fig:sphere1}
\end{figure}

For a choice of forcing parameters to be compatible with co-rotation, there must be an $\hat{F}_b$ that \textit{simultaneously} meets the conditions (\ref{eq:normeq}) and (\ref{eq:doteq})---that is to say the $\Omega$-curves and the $\theta$-curves must intersect.
Phrasing the problem in terms of the intersection between $\Omega$ curves and $\theta$ curves leads to an important observation.
If one of the $\Omega$ curves intersects one of the $\theta$ curves, then, generically, they must intersect again.
Additionally, by inversion symmetry, the antipodes of these to intersection points will also be intersection points.
Therefore, a choice of forcing parameters compatible with co-rotation actually gives rise to four co-rotating states.
An example of $\Omega$-curves and $\theta$-curves is illustrated in Fig.
\ref{fig:sphere1}.
This concludes our discussion of which forcing parameters give rise to co-rotation.

The next topic to be discussed is the stability of the co-rotating state.
Each of the four co-rotating states may have a different stability.
For there to be a globally stable co-rotating state, one and only one of these states must be stable.
We will show this to be true.

To this end, we will analyze the object's motion in the frame rotating with $\hat{F}$ at constant angular velocity $\vec{\Omega}$.
Note that in general, the object need not co-rotate with $\hat{F}$ and thus the object frame is in general distinct from the rotating frame.
We define $\bb{T}_I$ to be the twist matrix in this frame, so that $\bb{T}_I$ is defined by the following formula: \begin{equation}
 	\label{eq:TI}
 	\bb{T} \equiv \bb{R} \bb{T}_I \bb{R}^{-1}, \end{equation} where $\bb{R}(t) = \exp[\vec{\Omega}^\times t]$.
Notice that if the object is simply rotating with angular velocity $\vec{\Omega}$, then $\bb{T}_I$ is a constant.
We now find the equation of motion for $\mathbb{T}_I$.
Applying (\ref{eq:TI}) to (\ref{eq:diffeq}), we get, after some algebra, 
\begin{equation}
 	\label{eq:diffeqI}
 	\dot{\bb{T}}_I = [(\bb{T}_I \hat{F}_0 - \vec{\Omega})^\times ,\bb{T}_I], 
\end{equation}
where $\hat{F}_0 = \hat{F}(0)$.
This equation has the same form as the lab frame equation of motion (\ref{eq:diffeq}).
The only difference between these equations is that the angular velocity $\TT{} \hat{F}$ has been replaced by an apparent angular velocity $\bb{T}_I \hat{F}_0 - \vec{\Omega}$, as is appropriate for a frame rotating with angular velocity $\vec{\Omega}$.
Studying (\ref{eq:diffeqI}) reveals that there is a fixed point if the co-rotation condition, $\bb{T}_I \hat{F}_0 = \vec{\Omega}$, is satisfied.
This was the purpose of choosing the rotating frame.

We now study the stability of these fixed points.
To this end, we fix $\hat{F}_0$ and $\vec{\Omega}$ and assume that the twist matrix $\bb{T}_f$ is a fixed point (i.e.
$\bb{T}_f \hat{F}_0 = \vec{\Omega}$).
We will perform a linear stability analysis of this fixed point.
We may parameterize the departure of the twist matrix from $\bb{T}_f$ using an infinitessimal rotation vector $\vec{\eta}$.
We put \begin{equation}
 	\label{eq:tntf}
 	\bb{T}_I=\bb{T}_f+[\vec{\eta}^\times ,\bb{T}_f].
\end{equation}
Next we obtain the linearized equation of motion for $\vec{\eta}$.
We use the form (\ref{eq:tntf}) in (\ref{eq:diffeqI}) keeping terms linear in $\vec{\eta}$.
In the right hand side of (\ref{eq:diffeqI}), the first argument of the commutator represents the apparent angular velocity in the rotating frame; it must be first order in $\eta$. Since we are only working to first order, we may take only the zeroth order part of the right argument of the commutator. We find 
\begin{equation}
 \label{eq:etadiffeqsteps}
 [\dot{\vec{\eta}}^\times ,\bb{T}_f] =[([\vec{\eta}^\times ,\bb{T}_f]\hat{F}_0)^\times ,\bb{T}_f].
\end{equation}

Eq. (\ref{eq:etadiffeqsteps}) shows that $(\dot{\vec{\eta}} - [\vec{\eta}^\times ,\bb{T}_f]\hat{F}_0)^\times $ commutes with $\mathbb{T}_f$.
Unless the former is zero, this means that $\TT_f$ is unchanged upon an infinitesimal rotation in the $\dot{\vec{\eta}} - [\vec{\eta}^\times ,\bb{T}_f]\hat{F}_0 $ direction.
In this case, $\TT_f$ has an axis of symmetry.
Since we have restricted our attention to asymmetric $\mathbb{T}$'s, there can be no such axis; thus $\dot{\vec{\eta}}$ must equal $[\vec{\eta}^\times ,\bb{T}_f]\hat{F}_0$

Further manipulation yields
%\begin{widetext}
\begin{equation}
 \begin{aligned}
  \label{eq:etadiffeqsteps2}
  \dot{\vec{\eta}}&%
  =[\vec{\eta}^\times ,\bb{T}_f]\hat{F}_0 \\
  &%
  =\vec{\eta} \times \bb{T}_f\hat{F}_0 - \bb{T}_f (\vec{\eta} \times \hat{F}_0) \\
  &%
  =-\bb{T}_f\hat{F}_0 \times \vec{\eta} + \bb{T}_f (\hat{F}_0 \times \vec{\eta})\\
  &%
  =-(\bb{T}_f\hat{F}_0)^\times \vec{\eta} +\bb{T}_f \hat{F}_0^\times \vec{\eta},
 \end{aligned}
\end{equation}
and so we have the following differential equation for $\vec{\eta}$:
\begin{equation}
 	\label{eq:etadiffeq}
 	\dot{\vec{\eta}}=[\bb{T}_f \hat{F}_0^\times-(\bb{T}_f\hat{F}_0)^\times] \vec{\eta}.
\end{equation}
From this equation we find that the condition for the fixed point $\bb{T}_f$ to be locally stable is that all the eigenvalues of the ``stability matrix" $\bb{T}_f \hat{F}_0^\times-(\bb{T}_f\hat{F}_0)^\times$ have negative real part.

%\add[BM]{Ideally we would like to use these results to show that for any choice of forcing parameters, there is exactly one locally stable fixed point. However, this need not be the case. In some cases there may be no stable fixed points; in others there may be multiple. For this reason we must now restrict ourselves to a certain regime of forcing parameters, and show that in this case, there is exactly one stable fixed point.} \remove[BM]{Now that we have conditions for there to be fixed point solutions to }(\ref{eq:diffeqI})\remove[BM]{ for a given choice of $\Omega$ and $\theta$, and we are able to characterize the stability of these fixed points, we would like a way of being able to pick $\Omega$ and $\theta$ so that we are guaranteed a stable fixed point. }

We have seen above that a given rotating force gives in general four possible co-rotating orientations of the object.
Our goal of complete alignment would require that only one of these orientations survives after a long times.
While this does not happen for all values of forcing parameters, we will now show that there is a suitable choice of forcing parameters for which a single stable fixed point exists.

\begin{figure}
 \includegraphics[width=60mm]{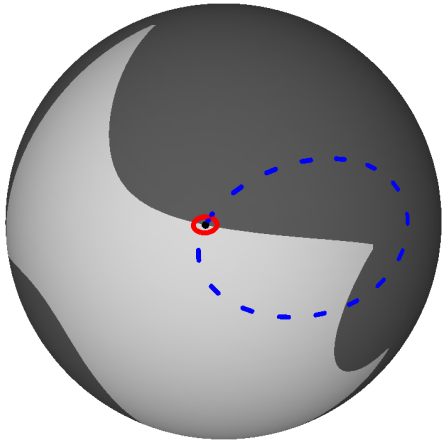}
 \caption{Sphere of $\hat{F}_b$'s for the case $\Omega=\lambda_3$ and small $\theta$.
  As in Fig.
  \ref{fig:sphere1}, the dashed curve is the $\Omega$-curve, the solid curve is the $\theta$-curve, and the black dot indicates the position of $\vec{v}_3$.
    The $\Omega$-curve passes through $\vec{v}_3$, since $\Omega=\lambda_3$, and the $\theta$-curve is a small loop around $\vec{v}_3$, since $\theta$ was chosen small. As seen in the figure, this means that these curves intersect in exactly two places. (The other two intersection points are the antipodes of the ones shown in the figure.)
  The light region is the ``stable region", where co-rotating states are locally stable.
  Only one of the visible intersections lies in the stable region; both of the fixed points that can't be seen are outside of the stable region, as explained in the text.}
 \label{fig:smalltheta}
\end{figure}

We consider the regime where $\theta$ is very close to $0$ and $\Omega =\lambda_3$.
Let's determine where the $\Omega$- and $\theta$- curves intersect in this case.
Since $\theta$ is so small, the $\theta$-curves are very small circles around $\pm \vec{v}_3$.
The $\Omega$ curves are a little more complicated.
Since we have chosen $\Omega$ to be the one generated by $\hat{F}_b=\vec{v}_3$, $\vec{v}_3$ is on the $\Omega$-curve.
Moreover, one readily verifies that the $\Omega$-curve must be \textit{smooth} at $\vec{v}_3$ as follows.
One can show from the form (\ref{eq:Tb}) of $\mathbb{T}$ given below that for the asymmetric objects considered here, $\vec{v}_3$ cannot be an eigenvector of $\mathbb{T}^T$, and thus it cannot be an eigenvector of $\mathbb{T}^T \mathbb{T}$.
Since the $\Omega$-curves are degenerate only at the eigenvectors of $\mathbb{T}^T \mathbb{T}$, the $\Omega$-curves must be smooth at $\vec{v}_3$ as claimed.
Thus, at $\vec{v}_3$ the $\Omega$-curve is almost straight on the scale of the $\theta$-curve.
Therefore, there will be exactly four intersection points, two near $\vec{v}_3$ and two near $-\vec{v}_3$.
This situation is shown in Fig. (\ref{fig:smalltheta}).

It may be illuminating to note the similarity between the force used in the forcing program proposed here, and the magnetic field used in nuclear magnetic resonance (NMR).
In NMR, there is a large DC magnetic field applied, causing the spins to have a response characterized by some angular velocity $\vec{\omega}$.
In the plane perpendicular to the DC magnetic field, a small (corresponding to $\theta \approx 0$) AC magnetic field is also applied with angular frequency $\vec{\Omega}=\vec{\omega}$.

Now, we will show that if the forcing parameters are chosen in this regime, only one fixed point is locally stable.
More specifically, we will show that there is only one fixed point at which all the eigenvalues of the stability matrix have negative real part.
To do this we will set $\Omega=\lambda_3$, and pick for the parameter $\theta$ some arbitrary, small value $\theta^*$.
We fix our attention on just the four fixed points resulting from this choice of forcing parameters.

First we will consider the two fixed points where $\hat{F}_b$ is near $-\vec{v}_3$.
Since these $\hat{F}_b$'s are so close to $-\vec{v}_3$, we would expect the behavior to be similar to the case where $\hat{F}_b$ \textit{is} $-\vec{v}_3$.
Therefore, let us study this latter case.
Since $-\vec{v}_3$ is an eigenvector, $\vec{\Omega}$ is parallel to $\hat{F}_b$, so $\theta=0$ and the force is constant.
We know from previous work \cite{Krapf09} that, at this fixed point, two directions are unstable while the third is neutrally stable.
%From this we infer that the stability matrix at this point has two eigenvalues with positive real part and one with zero real part.
%In fact, the third eigenvalue has to be exactly zero with corresponding eigenvector $-\vec{v}_3$, since rotations about this axis keep the object in a (rotating frame) fixed point.
%This concludes our discussion of the constant force fixed point; we turn our attention back to the two fixed points at $\theta=\theta^*$.
%They are only different from the constant force fixed point because their $\hat{F}_b$ is rotated slightly.
%From (\ref{eq:etadiffeq}), it is clear that a small change in the force can produce only a small change in the stability matrix, hence only a small change in the eigenvalues.
%Therefore the two eigenvalues of their stability matrix must have positive real part for sufficiently small $\theta$.
%This proves that these two fixed point are unstable.
The stability matrix varies smoothly with $\theta$ near $\theta = 0$.
Thus for a sufficiently small value $\theta^*$ of $\theta$, the unstable eigenvalues remain unstable, so that these fixed points are unstable.

We may perform a similar analysis for the two fixed points where $\hat{F}_b$ is near $+\vec{v}_3$.
Here, for $\theta=0$, there are two stable eigenvalues and a zero eigenvalue.
Therefore, the two $\theta=\theta^*$ fixed points must be stable in two directions, but overall stability requires that they also be stable in the third direction.

To determine the stability in the third direction, we will consider the $\theta=\theta^*$ fixed points as perturbations of the $\theta=0$ fixed point.
When $\theta$ is increased from 0, the co-rotating $\hat{F}_b$'s move outward from $\vec{v}_3$ along the $\Omega$-curve in opposite directions.
The stability matrix in the body frame is $\bb{T}_b \hat{F}_b^\times-(\bb{T}_b\hat{F}_b)^\times$; it is linear in $\hat{F}_b$.
Since the change in $\hat{F}_b$ is opposite for the two fixed points, the changes in the stability matrix, and, therefore, the eigenvalues are opposite.
The neutral eigenvalue either has opposite signs at the two fixed points or it remains zero for both.
All that is left to do then is demonstrate that the change in the eigenvalue is non-zero.

Our strategy is to assume this change is zero and obtain a contradiction.
First, we will choose a coordinate system, then we will write out the eigenvalue equation to lowest order in $\theta$.
We begin by picking our body frame coordinate axes, $\hat{x}$, $\hat{y}$, and $\hat{z}$, so that
\begin{equation}
 \label{eq:Tb}
 \mathbb{T}_b = 	\left[
 		\begin{array}{ccc}
  			t_{11}& t_{12}&
  0\\
  t_{21}&
  t_{22}&
  0\\
  0&
  b&
  \lambda_3
  		\end{array}
 	\right].
\end{equation}
Then the tangent direction of the $\Omega$-curve at $\vec{v}_3$ is the direction in the tangent space of the unit sphere which is perpendicular to $\mathbb{T}_b^T \mathbb{T}_b \vec{v}_3=(0,b \lambda_3, \lambda_3^2)$.
This direction is $\hat{x}$.
Working to first order in $\theta$, we write $\hat{F}_b^T = ( \theta ,0,1)$.
The third eigenvector of the stability matrix will have changed by some small amount $\vec{w} = (w_1,w_2,0)$ so that the eigenvector is now $\hat{z} + \vec{w}$, where $w_1$ and $w_2$ are small when $\theta$ is small.
Since we assume the eigenvalue is zero to lowest order, the eigenvalue equation satisfied by $\vec{w}$ reads: $(\mathbb{T}_b \hat{z}^{\times} - (\mathbb{T}_b \hat{z})^\times) \vec{w} + (\mathbb{T}_b \theta \hat{x}^\times - (\mathbb{T}_b \theta \hat{x})^\times) \hat{z} = 0$, where we have dropped terms higher than first order in $\theta$.
Notice that this equation is linear in $\theta$, $w_1$, and $w_2$, and so we may, by standard methods, determine if this equation has non-trivial solutions.
The result is that the equation has non-trivial solutions only when at least one of the two conditions are satisfied: (1) $t_{11} = \lambda_3$ and $t_{21}=0$ or (2) $b=0$.
The case (1) implies that $\lambda_3$ is an eigenvalue of $\mathbb{T}$ with multiplicity $2$, contradicting the fact that $\mathbb{T}$ has complex eigenvalues.
The case (2) implies that the object twist matrix has two-fold rotational symmetry about the $z$-axis, which we have assumed not to be the case.
We are forced to conclude that there is no solution to the eigenvalue equation if the eigenvalue is taken to be zero.
Evidently, then, the eigenvalue cannot remain zero to linear order.
This implies, as argued above, that there is only one stable fixed point.
This concludes our proof that there is one locally stable fixed point if the forcing parameters are chosen so that $\Omega=\lambda_3$ and $\theta$ is small.
In Section \ref{subsec:nRotF}, we will discuss numerical simulations showing that the fixed point is in general globally stable.

\section{Numerical Simulations}
\label{sec:numsim}
We carried out numerical simulations to test the results of applying the forcing programs outlined in Sections \ref{subsec:cstepfn} and \ref{subsec:cRotF} above.
In addition to confirming the behavior predicted in the previous section, the goals of the simulations are to get a sense of: (1) whether the decrease in entropy resulting from the step function forcing is indeed accompanied by complete alignment; (2) whether the condition of having $\psi$ monotonic is a necessary condition for achieving complete alignment; (3) whether the locally stable fixed point of the rotating force program is globally stable; and (4) whether alignment is possible outside of the small $\theta$ regime.
The results of our simulation will be presented in this section and the above four points will be discussed.

The simulations were done the following way: (A) we obtain a twist matrix, $\mathbb{T}$, by generating a $3 \times 3$ matrix with entries drawn uniformly from the interval $[-1,1]$; we reject any symmetric or non-aligning matrices to ensure that the matrix we obtain is axially aligning.
(B) We subject the twist matrix to a constant force along the $z$-axis so that its real eigenvector is aligned with that direction.
This is done by solving the differential equation (\ref{eq:diffeq}) with $\mathbb{T}$ as initial data.
(C) 500 angles are drawn uniformly from the interval $[0,2\pi]$, and new twist matrices are then formed by rotating the twist matrix found in step (B) by the random angles.
These 500 $\TT$'s form our axially aligned ensemble.
(D) These new twist matrices are then subjected to a forcing program, which again is done by numerically solving (\ref{eq:diffeq}) with the appropriate initial conditions.
This will be discussed in more detail for each forcing program.
(E) The orientation of each matrix about the $z$-axis is determined.
(In the case of the rotating force, a constant force along the $z$-axis is first applied to ensure that the twist matrices are at least axially aligned.) (F) This process is repeated for many different random initial twist matrices in order to explore the range of different possible outcomes.
In the following two subsections we will discuss step (D) of the simulations and present results both forcing programs.

\subsection{Step Function Forcing}
\label{subsec:nstepfn}
\begin{figure*}
 \resizebox{150mm}{!}{\includegraphics{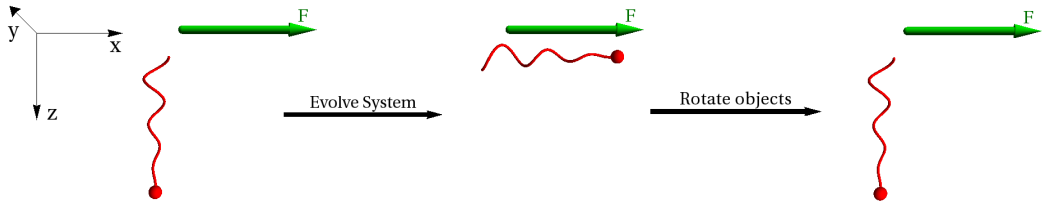}}
 \caption{Above is a schematic illustration of one step function iteration.
  On the far left of the figure is the lab coordinate axes.
  There are three stages shown, and for each stage we show the force vector in green, and a representative object of the ensemble in red (in reality there are many objects in the ensemble, but only one is depicted for ease of viewing).
  At the beginning of an iteration, the ensemble is axially aligned with the $z$-axis, and a force is applied along the $x$-axis.
  After the ensemble is given some fixed time to evolve, the ensemble becomes axially aligned with the $x$-axis.
  After this, the ensemble is allowed to rotate for a random amount of time.
  The next step in the simulation is to rotate the objects by hand so that they become aligned with the $z$-axis again.
  After this step, we are ready to begin the next iteration.}
 \label{fig:stepiterfig}
\end{figure*}

As mentioned in the previous paragraph, our aim is to test if complete alignment occurs, since this is not guaranteed by the decreasing of entropy.
We also test whether it is possible to effect complete alignment even when $\psi(\phi)$ is non-monotonic.
To study the behavior, we use the program described in the previous paragraph, where step (D) is carried out in the following three stages: first the objects, formed by step (C), which are aligned with the $z$-axis, are subjected to a constant force along the $x$-axis until they are aligned to a resolution of $10$ milliradians.
This stage is the physical manifestation of applying $\psi$.
Second, they are all allowed to rotate for an additional time randomly drawn from a uniform distribution ranging from zero to the period of rotation.
This stage applies the shift.
Finally an explicit $90^{\circ}$ rotation about the $y$-axis is applied to each $\mathbb{T}$ matrix, so that they are again aligned with the $z$-axis.
We then repeat the process starting from stage 1.
Each repetition is a ``step function iteration." The step function iteration is illustrated in Fig.
\ref{fig:stepiterfig}.
We subject the system to as many step function iterations as desired (which in our simulations was 100).
After each iteration, the orientation of each object is determined, and the entropy of the ensemble is found using a nearest neighbor estimate \cite{Kozachenko87}.

\begin{figure*}
 \resizebox{150mm}{!}{\includegraphics{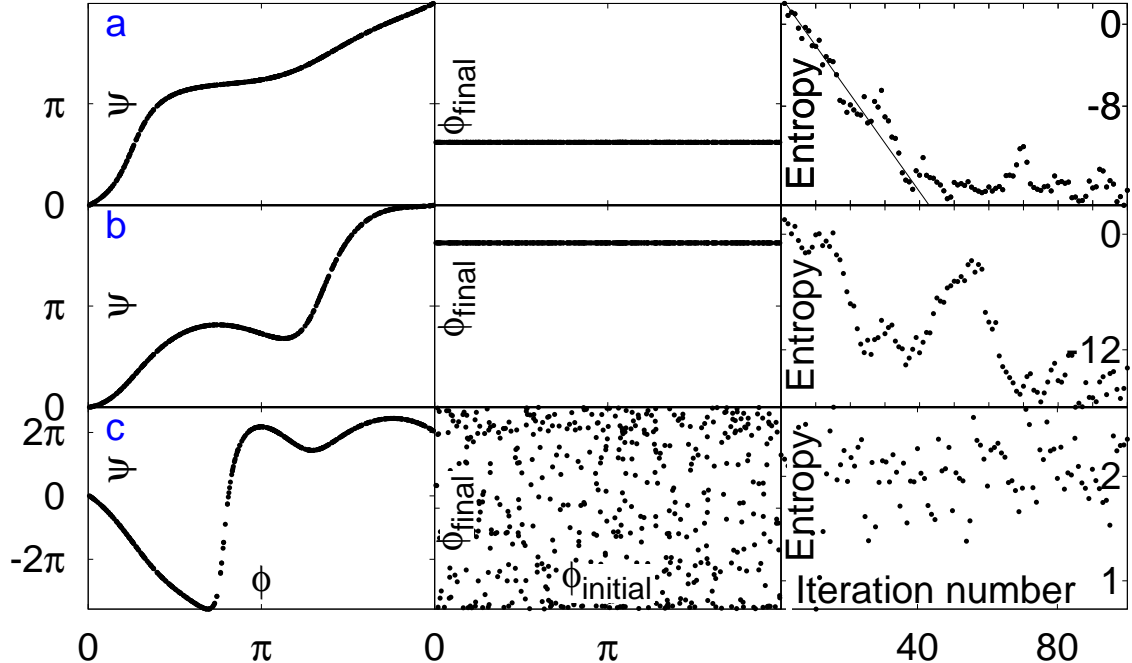}}
 \caption{Results of simulating the alternating forcing program for three different twist matrices.
  Each row of plots corresponds to a different twist matrix.
  In each row, the first plot shows $\psi(\phi)$ vs $\phi$; the second plot shows $\phi_{\mathrm{final}}$, which is the orientation resulting from 100 step function iterations, vs $\phi_{\mathrm{initial}}$; and the third plot shows the entropy as a function of iteration number.
  The line in the first row indicates the decrease of entropy in bits predicted by (\ref{eq:dhpsi})}
 \label{fig:allplots}
\end{figure*}

Some of our results are shown in Fig.
\ref{fig:allplots}.
Each row of plots corresponds to a different twist matrix.
The first row is associated with a twist matrix whose $\psi$ function is monotonic.
From left to right, the plots in this (and every other) row are: $\psi(\phi)$ vs $\phi$, the final orientation after all 100 iterations vs initial orientation, and the entropy (in bits) as a function of iteration number.
We can see from the second plot in the first row that the final orientations all share a common value; i.e., the objects have aligned.
Thus we did not run into the problem of having multiple final orientations, even though this would have been consistent with our proof of indefinite decrease in entropy.
In the third plot of the first row, there is a roughly linear trend between entropy and iteration number, consistent with our prediction in (\ref{eq:dhpsi}).
The final entropy, which is approximately $-16$ bits, can be attributed to numerical error.
In our simulations, each object that had a monotonic $\psi$ exhibited behavior similar to this example.
 In particular, we found that complete alignment was attained.

The next row shows that ordering can occur even with a non-monotonic $\psi$.
We see in the second plot that although $\psi$ is non-monotonic, all the objects have the same final orientation.
The entropy decrease is more erratic in this case.
However, not all twist matrices that we examined gave alignment.
For example, the twist matrix of the third row resulted in a random distribution of final orientations.
Correspondingly, we see that the entropy remains near its initial value.

\subsection{Rotating Force}
\label{subsec:nRotF}
In the case of the rotating force, our goals were to test if the stable fixed point is always globally stable when $\theta$ is small, and to determine the behavior for larger $\theta$, where we have not even shown a stable fixed points exists.
To test global stability for small $\theta$, we performed step (D) of the simulation by starting with an axially aligned ensemble, and applying a sedimenting force of the form (\ref{eq:fxy}) to an ensemble of objects, with $\theta$ small, and $\Omega = \lambda_3$.
To measure the degree of alignment of the ensemble at any point during the forcing, we would make a copy of the ensemble, and apply a constant force along the $z$-axis so the ensemble would at least be axially aligned.
Then the orientation of each object is determined, and the standard deviation of the set of orientations is computed.
Then relative orientations are determined and the ensemble is considered aligned if this standard deviation is less than 0.5 milliradians.
Having done this simulation on 1000 different randomly generated twist matrices, we found that alignment is always achieved if $\theta$ is chosen to be sufficiently small and the system is evolved for a sufficiently long time.
The typical value of $\theta$ was about $0.07$ radians, but, as we will show, $\theta$ could have been chosen to be larger in many cases.
The typical amount of time required for alignment is about 500 periods of the forcing.

However, it was occasionally the case that alignment took a very long time.
The twist matrices which took longest to align were ones where $\vec{v}_3$ was very near one of the eigenvectors, call it $\vec{w}$, of $\mathbb{T}^T \mathbb{T}$.
%(\ref{eq:Tb})
Before determining why this leads to slow alignment, we identify experimental behavior connected with nearly coinciding eigenvectors.
In the coordinate system of (\ref{eq:Tb}), $\mathbb{T}^T \mathbb{T} \vec{v}_3 = \lambda_3 ( 0,b,\lambda_3)$.
Thus we find that $\vec{v}_3$ is an eigenvector of $\mathbb{T}^T \mathbb{T}$ only if either $\lambda_3=0$ or $b=0$.
Therefore, if $\vec{v}_3$ is nearly and eigenvector of $\mathbb{T}^T \mathbb{T}$, either $\lambda_3$ or $b$ must be small.
The meaning of a small $\lambda_3$ is that the object has a slow response to a constant force, and thus the rotating force is required to rotate very slowly.
The meaning of $b=0$, as can be verified by direct computation, is that $\hat{z}$ is a two-fold symmetry axis of $\mathbb{T}$.
Therefore, the problem twist matrices are ones where either the force is required to rotate very slowly or the twist matrix was nearly two-fold symmetric.

The small difference between $\vec{v}_3$ and $\vec{w}$ leads to slow alignment for two reasons to be explained below.
First, $\theta$ must be small in order to satisfy both the $\Omega$ and the $\theta$ constraints.
Second, a small $\theta$ entails slow alignment.

We first explain why $\theta$ must be small when the eigenvectors nearly coincide.
On the one hand, the $\mathbb{T}^T \mathbb{T}$ close to $\vec{v}_3$ may correspond to an \textit{extremal} eigenvalue of $\mathbb{T}^T \mathbb{T}$.
In this case, one $\Omega$-curve is closely localized around $\vec{w}$.
Thus if $\theta$ is chosen too large, the $\theta$-curve will completely circumscribe the $\Omega$ curve, and these curves will not intersect.
Thus $\theta$ must be small if $\vec{v}_3$ is close to $\vec{w}$ in this case.
On the other hand, $\vec{w}$ may correspond to the intermediate eigenvalue.
Then the two solution curves of (\ref{eq:normeq}) are almost touching at $\vec{v}_3$.
Therefore, if $\theta$ is chosen too large, the $\theta$ curve will intersect both $\Omega$ curves, and there will be a total of $8$ fixed points, possibly two of them stable.
Thus $\theta$ must be small in this case as well.

The smallness of $\theta$ slows the alignment because as $\theta$ goes to $0$, the third eigenvalue of the stability matrix also must go continuously to zero.
Thus the time required for alignment becomes very long, and the numerical simulation is difficult.
Of course this problem is relevant to physical applications as well as numerical simulations.

Next we consider the behavior outside of the small $\theta$ regime.
To explore the full range of $\theta$ by simulation, multiple copies of the initial ensemble were created, each one subjected to a different forcing program.
The $\theta$'s of these forcing programs were evenly distributed between $0$ and $\pi$, but as before we set $\Omega = \lambda_3$.

We characterize how ``well" an object aligns as a function of $\theta$ in the following sense.
If the ensemble is sufficiently converged, then the system is well described by a linear approximation, and the standard deviation of the orientations must decay exponentially with time.
%We found this exponential decay rate in each case where alignment occurred. 
We will use this decay rate to quantify how well the object aligns.
%This decay as a measure of how quickly the ensemble aligns. 
We know that this decay rate is small when $\theta$ is small, but we also have reason to believe that alignment is impossible if $\theta$ is too large.
Thus there is some intermediate optimal $\theta$, and studying the alignment rate vs.
$\theta$ should give an indication of what the optimal $\theta$ tends to be.

\begin{figure*}
 \begin{tabular}{ccc}
  \includegraphics[scale=.5]{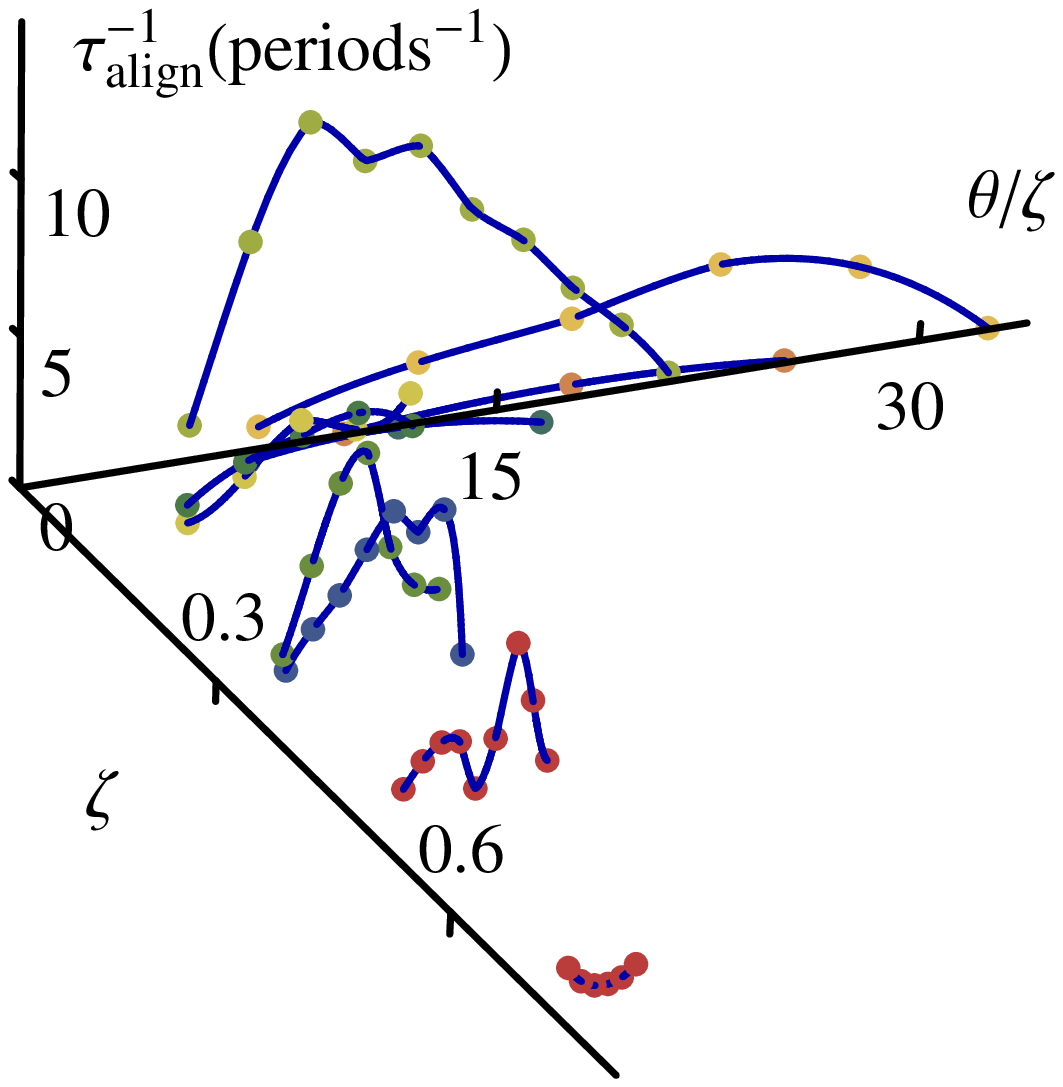} &
  \includegraphics[scale=.5]{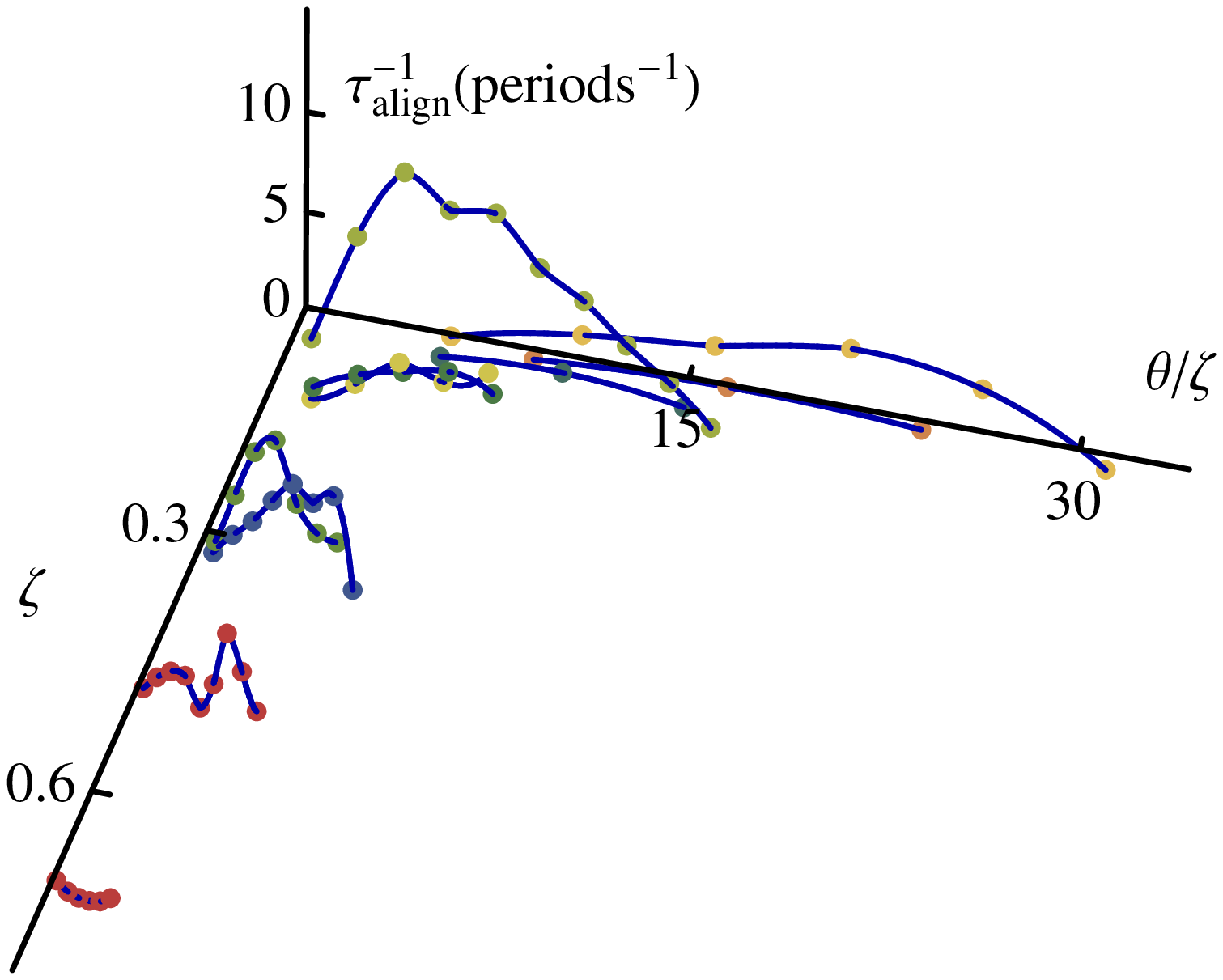}
 \end{tabular}
 \caption{ Rates of full alignment of a sample of ten randomly generated twist matrices subjected to rotating forcing programs of various $\theta$ as described in the text.
  For clarity two views of the three-dimensional plot are shown, and color is used to distinguish points corresponding to different twist matrices.
  The vertical axis shows the rate of alignment $\tau^{-1}$, expressed in units of the respective object's natural rotation period under constant force.
  The back-to-front axis represents the angle $\zeta$ between the stable eigenvector $\vec{v}_3$ of $\mathbb{T}$ and the nearest eigenvector of $\mathbb{T}^T \mathbb{T}$.
  The left-to-right axis is the angle $\theta$ normalized to this $\zeta$.
  Left-to-right curves show the effect of increasing the angle $\theta$ between the force and the rotation axis for a given twist matrix.
  The interpolating curves are guides for the eye.
  Points corresponding to values of $\theta$ that did not lead to alignment are omitted from the plot.
 }
 \label{fig:largerTheta}
\end{figure*}

Fig \ref{fig:largerTheta} shows simulation results for $10$ different twist matrices.
We parameterize each twist matrix by the angle $\zeta$ between $\vec{v}_3$ and the closest eigenvector of $\mathbb{T}^T \mathbb{T}$.
We argued above that if the $\zeta$ angle for a particular $\mathbb{T}$ is small, then $\theta$ must be small in order for alignment to occur, and the alignment must be slow.
Although the alignment for small $\zeta$ is typically slow, we find that for any $\zeta$---small \textit{or} large---the optimal $\theta$ is typically larger than $\zeta$, showing that not only is our method robust to picking a large value for $\theta$, but choosing a large $\theta$ is often beneficial.
Remarkably, the optimal alignment rate can be many multiples of the rotational frequency, although the total time required for alignment is even then typically a few rotational periods, since alignment in the non-linear regime is slower. Additionally, we found that when alignment did occur, it was always accompanied by co-rotation.

\section{Discussion}
\label{sec:disc}
\subsection{Assumptions}
\label{subsec:Assumptions}
In the above derivations, we have made many assumptions about the nature of the medium, the nature of the forcing, and the nature of the sedimenting object.
We will now take inventory of these assumptions and discuss their validity.
One of the fundamental assumptions we made in our approach to this problem was to neglect interactions between the different objects in the ensemble, thereby showing that the orientational alignment demonstrated here does not require these interactions.
Thus the underlying mechanism is different from the one encountered with weakly coupled dynamical systems such as a row of similar mechanical clocks mounted on a wall \cite{Vollgraff50}.
However, in practice there is a coupling: a given object in the suspension experiences forces and torques owing to the flow emanating from nearby sedimenting objects.
These interactions may enhance or diminish the phase locking effects shown above.
In any case, the interactions can be reduced by diluting the suspension.

Being acted upon by a force, the object will move.
We assumed, as would be appropriate for objects of colloidal length scales, that the resulting motion is characterized by a small Reynolds number.
Typical objects up to the size of a millimeter easily satisfy this criterion.
Experiments \cite{Krapf09} up to Reynolds numbers of several hundred showed the same sedimentation behavior as for Reynolds numbers below 1.
However, inertial effects at higher Reynolds numbers may well alter our results.
The assumptions listed so far allowed us to state that the angular velocity of an object at any point in time depends only on the force applied to it at that moment, and it depends on the force linearly through the twist matrix.

In writing (\ref{eq:oTF}), we have regarded the twist matrix in the body frame as constant.
Since this matrix depends only on the shape and mass (or charge if the forcing is electrophoretic) distribution of the object, we have essentially assumed that the shape of the object does not change with time, or, in other words, the object is rigid.
This assumption is well satisfied for many objects.
Still, any given object may be deformed by the hydrodynamic stresses.
A sufficiently large deformation creates corrections to the linear response we have treated here.

We have also assumed that the objects in the ensemble have identical twist matrices.
In practice any small differences in the twist matrices leads to misalignment.
In the case of the step function forcing, one might worry that small changes is the twist matrix might lead to a vastly different aligned state after many iterations.
However, the ability of our simulations to work despite representing the twist matrix to only machine precision indicates that this is not a problem.
In the case of a rotating force, whether or not the $\Omega$-curves and $\theta$-curves intersect should be insensitive to small changes in the object shape.
Therefore, we would still expect an ensemble of slightly different objects to co-rotate with each other, given appropriately chosen forcing parameters.

We have supposed that the object rotates deterministically in response to $\vec{F}$.
In practice, thermal fluctuations add random rotations to this response.
Our approximation is valid provided that the object size is sufficiently large, since the size of the thermal fluctuations decreases with increasing object size.
The minimum object size will be estimated in Section \ref{subsec:exr}.

Another assumption we made was that the twist matrix was ``axially aligning", with only one real eigenvalue.
The motivation for this assumption was that under a constant force, these object become axially aligned.
If this assumption is not in force, then there are two different axes in the body that may align with a constant force, so there is no way to characterize the transient by a single function $\psi$ and the formalism used to demonstrate alignment completely breaks down.
However, the rotating force may still align, and this is a possible direction of future research.

A further assumption made above was to consider generic objects with no rotational symmetry.
Since $\mathbb{T}$ is axially aligning, it necessarily has a preferred axis which must be preserved by any symmetry.
Therefore, the only symmetries $\mathbb{T}$ can possess are discrete $n$-fold rotations about this axis or complete circular symmetry about the axis.
With complete rotational symmetry, further alignment is impossible since the axially aligned $\mathbb{T}$'s are already identical.
The only $n$-fold symmetry to consider is $2$-fold symmetry since higher symmetries are impossible.
\footnote{ A rotational symmetry of a tensor must simultaneously be a symmetry of both the symmetric and antisymmetric parts.
 The antisymmetric part has the form $\vec{u}^\times$ for some $\vec{u}$.
 The only symmetries the antisymmetric part has, then, is rotations about $\vec{u}$.
 The symmetric part defines three eigenvalues $\lambda_1$, $\lambda_2$, and $\lambda_3$, with eigenvectors $\vec{v}_1$, $\vec{v}_2$, and $\vec{v}_3$.
 If the eigenvalues are distinct, then the eigenvectors are uniquely defined up to a sign.
 Any rotational symmetry in the distinct eigenvalues case must then be a $\pi$ rotation about one of the eigenvectors, and no higher symmetry than two-fold rotations is possible.
 If two eigenvalues are the same, then the symmetric part has circular symmetry about the non-degenerate eigenvector.
 In any event, rotational symmetries are only possible when $\vec{u}$ is an eigenvector of the symmetric part, but then the only symmetries possible are two-fold symmetry and circular symmetry.
}
In the case of a two-fold symmetry, the step function forcing results are easily extended: where previously an objects orientation was represented by a number $\phi$ in the interval from $0$ to $2 \pi$, here we need only consider orientations in the interval from $0$ to $\pi$.
The function $\psi$ maps this interval into itself.
Since there is no essential difference between this interval and the full interval from $0$ to $2 \pi$, the proof of alignment goes through unmodified.
However the case of rotating force is not so straightforward.
Here $\vec{v}_3$ is necessarily an eigenvector of $\mathbb{T}^T \mathbb{T}$, and we are unable to use the same arguments that we did in the case of a non-symmetric \TT{}.
This is a direction for future work.

In addition to these global assumptions we made further restrictions particular to either forcing method.
For the step-function method, we assumed that the function $\psi$ was monotonic.
We found this could be easily attained by making the rocking angle $\theta$ small enough.
In the case of the rotating force, we found that a good choice of parameters for the rotating force program was similar to what is done in nuclear magnetic resonance: we have the DC component of the force be much larger that the AC component, and we have the AC component rotate at the angular velocity given by the response of the object to the DC component.
Although making $\theta$ small should be easy to do experimentally, it may not be easy to set $\Omega=\lambda_3$, since $\lambda_3$ may not even be known.
One way to surmount this problem is to gradually increase the frequency with time.
As the frequency is swept through $\lambda_3$, the ensemble will become completely aligned.
Once the frequency gets beyond $\lambda_3$, the objects will evolve in a similar way since they are in a similar orientation, so the ensemble will remain aligned.

\subsection{Thermal Diffusion}
\label{subsec:exr}
%Having described the assumptions necessary for alignment, we will now give an example of a system that meets these requirements. This system could be used to experimentally test the aligning power of the two forcing programs listed above.

In Section \ref{subsec:Assumptions} we noted that if the object was not large enough, thermal rotational diffusion would make alignment impossible.
Thus there is a minimum object size required for alignment to occur.
Our goal in this subsection is to estimate this minimum size.
To do this, we will first estimate the time for an object to lose its orientation as a function of size.
Then we will estimate the time it takes for an ensemble to become aligned as a function of size in two scenarios.
The first scenario is when the physical origin of the aligning force is centrifugation, and the second scenario is when the force is due to electrophoresis.
First, we calculate the rate at which an object loses its orientation.
We will call this the decoherence rate $\Gamma$.
Denoting the typical angle an object has rotated from the aligned state as $\Delta \phi$, the criterion for the ensemble to be disordered is that $\Delta \phi \approx 1$.
In order to estimate diffusion, 
%we will assume that the object's angular speed is just it's speed divided by its hydrodynamic radius, i. e.,  $\Delta \phi = \frac{\Delta R}{R_H}$. The translational diffusion of the object is described by $(\Delta R)^2 = 2 D t$, where the diffusion constant $D=\frac{kT}{6 \pi \eta R_H}$. Plugging these in, we find the decoherence rate, $\Gamma=\frac{kT}{3 \pi \eta R^3_H}$.
we approximate the object as a sphere whose diameter $2R$ is the span of the object \cite{Witten10}.
The rotational diffusion constant $D_r$ for such a sphere at temperature $T$ is given by $D_r = k_B T/(8 \pi \eta R^3)$, where $k_B$ is Boltzmann's constant.
Then over short times $t$, the mean-squared angular displacement $\Delta \phi$ is given by $(\Delta \phi)^2 = 2 D_r t$.
This formula gives a decoherence rate $\Gamma \approx 2 D_r = \frac{k_B T}{4 \pi \eta R^3}$.

Next we will compare the decoherence rate to the rate at which the objects become aligned.
To do this, we must first find the typical angular velocity of an object.
First we consider the case where a centrifuge is providing the sedimenting force.
the centrifugal force is $F=\rho_b a_c V$, where $\rho_b$ is the absolute value of the object's buoyant density (the difference between the objects density and the density of the solvent); $a_c$ is the centrifugal acceleration felt by the object; and $V$ is the volume of solvent displaced by the object.
It is convenient to relate $V$ to the typical radius $R$ using a parameter $\alpha$ representing how much of the pervaded volume is occupied by the object: $ V=\frac{4 \pi}{3} \alpha R^3 $.
To estimate the angular speed, we observe that, by dimensional analysis, $\omega ~ v/R$ so that $\omega = \beta \frac{F}{6 \pi \eta R^2}$.
The constant of proportionality $\beta$ specifies the degree of rotation-translation coupling and depends on the shape of the object.
In our prior simulations of physical shapes, we found that $\beta$ is typically of order $0.01$ \cite{Moths13}.

The expression for the angular velocity becomes $\omega = \frac{2 \alpha \beta \rho_b a_c R}{9 \eta}$.
Now that we have found the angular velocity we will estimate the alignment rate.
As indicated by the large alignment rates in Fig.
\ref{fig:largerTheta}, objects typically align in just a few periods of rotation.
Therefore we will define the alignment time to be $ t_\mathrm{a} = \frac{1}{\gamma \omega}$ with $\gamma = 0.1$.
Setting the product of the alignment time with the decoherence rate equal to unity, we obtain the following lower limit on the radius: $R > \sqrt[4]{\frac{3 k_B T}{2 \alpha \beta \gamma \pi \rho_b a_c}}$.
Putting $\alpha=1$, $T=300\mathrm{\ K}$, $\rho_b=100 \mathrm{\ kg/m^3}$, and $a_c=10^5 \mathrm{\ m/s^2}$, we find $R>0.7 \mathrm{\ \mu m}$.

As noted in Sec.
\ref{subsec:rotres}, our alignment mechanism depends on a linear coupling between a vector perturbation and the rotation of the objects.
Since asymmetric objects rotate in response to an electric field \cite{Long98}, our alignment methods may be implemented via electrophoresis.
When an electric field $\vec{E}$ is applied, an object will move with a velocity $\vec{v}=\mu \vec{E}$, where the proportionality constant $\mu$ is called the electrophoretic mobility.
The mobility depends linearly on the zeta potential $\zeta$.
Specifically $\mu = \frac{\epsilon \zeta}{\eta}$ where $\epsilon$ is the permittivity of the solvent.
Assuming, as we did in the case of sedimentation, that $\omega = \beta v/R$ and $t_\mathrm{a} = \frac{1}{\gamma \omega}$, and again requiring that $\Gamma t_\mathrm{a}>1$, we get the following lower bound on the radius: $ R > \sqrt{\frac{k_B T}{4 \pi \beta \gamma \eta E \zeta}} $.
To obtain a numerical value for $R$, we assume the strength of the applied electric field is $10^4\ \mathrm{V/m}$, the $\zeta$ potential is $15 \textrm{\ mV}$ (as is necessary to stabilize the colloidal suspension \cite{Riddick68}), and the relative permittivity of the solvent is $80$, as with water.
Assuming the same values as before for other constants, we find the minimum radius is $2 \mathrm{\ \mu m}$.

In either case, we have found a lower limit for the size of the object, as measured by the hydrodynamic radius, of order $1 \mathrm{\ \mu m}$.

\subsection{Applications}
\label{subsec:applications}

Polymer science \cite{Peters09}, molecular biology \cite{Seemans03} colloid science \cite{Wang12} and beam lithography \cite{folch12} have made it possible to produce identical asymmetric micron-scale objects in large quantities, e.g., pollen grains \cite{Katifori10}.
These objects often occur as dispersions in a liquid.
In current practice, the objects interact individually and stochastically with their surroundings, so that their relative orientations are not important.
However, the oriented samples envisaged here offer the prospect of coherent responses to their environment.
They move together in response to forces, or chemical gradients.
Thus a group of objects in part of the sample could be steered to migrate together into a desired region.
The oriented objects scatter waves in the same way.
Thus the determination of their structure via x-ray diffraction is simplified by reducing the orientational randomness.
Any anisotropic emissions such as fluorescent radiation or Soret currents are also enhanced by the common alignment of the objects.
Thus the possibility of tandem control of the orientation of the objects in a sample suggests new ways for characterizing them and manipulating them.

\section{Conclusion}
\label{sec:conclusion}

These possible applications give only a suggestion of the potential uses of programmed forcing of asymmetric colloidal objects.
Their value in practice will depend on the quantitative impact of degrading effects like thermal fluctuations and object variability mentioned above.
Another limitation comes from the twist matrices encountered in the actual objects to be manipulated.
We have seen above that these matrices can be controlled to a degree, but the actual scope for such control remains to be explored.
Our work to improve this understanding is in progress.
Conceptually, this work broadens our notions of how random objects can be organized by programmed external fields.
In the quantum world, the value of programmed magnetic and electric fields has been well appreciated as a means for organizing the disordered quantum states of atoms and molecules.
This work shows that similar means can be used to organize colloidal objects.
There is every reason to anticipate further extensions.

\section{Acknowledgments}
The authors are indebted to Michael Solomon for his suggestion that this work is applicable to electrophoresis.
We also thank Nathan Krapf for his guidance.
This work was supported in part by the National Science Foundation's MRSEC Program under Award No.
DMR-0820054.

\bibliography{../references/savedrecs}
\end{document}